\documentclass[prl,aps,showpacs,onecolumn]{revtex4}%
\usepackage{graphicx}
\usepackage{bm}%
\usepackage{amsmath}%
\usepackage{amsfonts}%
\usepackage{amssymb}
\providecommand{\U}[1]{\protect\rule{.1in}{.1in}}
\begin{document}
\title{ Dirac Particles in a Gravitational Field}
\author{Pierre Gosselin$^{1}$ and Herv\'{e} Mohrbach$^{2}$}
\affiliation{$^{1}$Institut Fourier, UMR 5582 CNRS-UJF, UFR de Math\'{e}matiques,
Universit\'{e} Grenoble I, BP74, 38402 Saint Martin d'H\`{e}res, Cedex, France }
\affiliation{$^{2}$Groupe BioPhysStat, ICPMB-FR CNRS 2843, Universit\'{e} Paul
Verlaine-Metz, 57078 Metz Cedex 3, France}

\begin{abstract}
The semiclassical approximation for the Hamiltonian of Dirac particles
interacting with an arbitrary gravitational field is investigated. The time
dependence of the metrics leads to new contributions to the in-band energy
operator in comparison to previous works on the static case. In particular we
find a new coupling term between the linear momentum and the spin, as well as
couplings which contribute to the breaking of the particle - antiparticle symmetry.

\end{abstract}
\maketitle

\section{Introduction}

In this paper we consider the theory of Dirac fermions in an arbitrary curved
space-time in the Hamiltonian formulation. To reveal the physical content of
the theory it is necessary to perform the diagonalization of the Hamiltonian
uncoupling the positive and the negative energy states. For a fermion
interacting with an electromagnetic field the Foldy-Wouthuysen (FW)
transformation based on an approximate scheme valid in the non relativistic
limit is often used \cite{FW}. This same method was also applied in all the
previous studies of Dirac fermions in a gravitational field \cite{OBUKHOV}.
Here instead, we will consider the fully relativistic regime but in a
semiclassical approximation for which the de Broglie wave length of the
fermion must be much smaller that the characteristic size of the
inhomogeneities of the external field. A recent semiclassical FW-like
transformation used for Dirac particle in a strong electromagnetic field could
be adapted to the gravitational problem \cite{SILENKO}. But instead we will
use another method developed by the authors which essentially differs from the
FW. This method allows us to find the diagonal representation of any kind of
matrix valued quantum Hamiltonian as a series expansion in the Planck
constant. Here we will directly apply the general formula obtained at the
semiclassical (first order in the Planck constant) limit to the case of Dirac
fermions in an arbitrary curved spacetime. This is an extension of previous
papers where massless and massive particles in a static gravitational fields
were treated. The extension to time dependent metrics turns out to be
non-trivial and leads to new coupling terms in the in-band energy operator
which break the particle-antiparticle symmetry.

\section{Electron in a Gravitational Field}

A one half spinning particle of mass $m$ coupled to an arbitrary gravitational
field is described by $4-$spinor field $\psi$ satisfying the covariant Dirac
equation
\begin{equation}
(i\hbar\gamma^{\alpha}D_{\alpha}-m)\psi=0\label{Dg}%
\end{equation}
where we use $c=1$, but keep explicit the Planck constant $\hbar$ and
$\alpha=0,1,2,3.$

The covariant spinor derivative is defined as $D_{\alpha}=h_{\alpha}^{i}D_{i}$
with $D_{i}=\partial_{i}-\frac{1}{4}\left[  \gamma^{\alpha},\gamma^{\beta
}\right]  \Gamma_{i}^{\alpha\beta}$. The matrices $\gamma^{\alpha}$ are the
usual Dirac matrices, $h_{\alpha}^{i}$ are the the orthonormal vierbein and
$\Gamma_{i}^{\alpha\beta}$ the spin connection components.

Rewriting Eq. $\left(  \ref{Dg}\right)  $ under the Schr\"{o}dinger form
\[
i\hbar\frac{\partial\psi}{\partial t}=H\psi
\]
we obtain the following Hamiltonian
\begin{equation}
H=g_{00}h_{\beta}^{0}\gamma^{\beta}\left(  \gamma^{\alpha}\bar{P}_{\alpha
}+m\right)  +\frac{\hbar}{4}\varepsilon_{\varrho\beta\gamma}\Gamma
_{0}^{\varrho\beta}\mathbf{\Sigma}^{\gamma}+i\frac{\hbar}{4}\Gamma_{0}%
^{0\beta}\alpha_{\beta}\label{Ham1}%
\end{equation}
where we introduced the notation for the pseudo-momentum $\bar{P}_{\alpha
}\mathbf{=}h_{\alpha}^{i}(P_{i}+\frac{\hbar}{4}\varepsilon_{\varrho\beta
\gamma}\Gamma_{i}^{\varrho\beta}\mathbf{\Sigma}^{\gamma})$ with the spin
matrix $\mathbf{\Sigma}^{\gamma}$, satisfying the relation $\varepsilon
^{\gamma\alpha\beta}\mathbf{\Sigma}_{\gamma}=\frac{i}{8}(\gamma^{\alpha}%
\gamma^{\beta}-\gamma^{\beta}\gamma^{\alpha}).$ We use the conventions of
Bjorken and Drell \cite{Bjorken} for the Dirac matrices $\gamma^{\alpha},$ and
$\alpha^{\beta}.$

Surprisingly, Eq. $\left(  \ref{Ham1}\right)  $ turns out to be non-hermitian
for a time dependent metric. We then follow the approach of Leclerc
\cite{Leclerc} who showed that one must add the term $\frac{i}{2}\partial
_{t}\ln\left(  -gg^{00}\right)  $ to make it Hermitian. It will be shown later
on, that the presence of this term is also necessary for the diagonalization
procedure to work.

Therefore the Hamiltonian considered in the following will be%

\begin{equation}
\hat{H}=g_{00}h_{\beta}^{0}\gamma^{\beta}\left(  \gamma^{\alpha}\bar
{P}_{\alpha}+m\right)  +\frac{\hbar}{4}\varepsilon_{\varrho\beta\gamma}%
\Gamma_{0}^{\varrho\beta}\mathbf{\Sigma}^{\gamma}+i\frac{\hbar}{4}\Gamma
_{0}^{0\beta}\alpha^{\beta}+\frac{i}{2}\partial_{t}\ln\left(  -gg^{00}\right)
\label{H1}%
\end{equation}
The goal of this paper is therefore to diagonalise Eq. $\left(  \ref{H1}%
\right)  $ to first order in $\hbar$. But before embarking into this, we need
first to discuss the definitions and properties of the scalar product in a
curved space-time.

\subsection{Scalar product.}

As said before the Hamiltonian $\hat{H}$ is Hermitian. However, the notion of
hermiticity is here defined with respect to a scalar product in curved space
\cite{Leclerc}, namely :
\begin{equation}
\left\langle \psi_{1}(t)\mid\psi_{2}\left(  t\right)  \right\rangle
_{U(t)}=\int\psi_{1}^{+}\sqrt{-g}h_{\beta}^{0}\gamma^{\beta}\gamma^{0}\psi
_{2}d^{3}x=\int\psi_{1}^{+}U\psi_{2}d^{3}x\label{SP}%
\end{equation}
where we introduced the notation $U=\sqrt{-g}h_{\mu}^{0}\gamma^{\mu}\gamma
^{0}$ and $\psi_{1}$, $\psi_{2}$ two spinors. Therefore an operator $O$ is
hermitian for $\left\langle \mid\right\rangle _{U(t)}$ defined in Eq. $\left(
\ref{SP}\right)  $, if
\begin{equation}
\int\psi_{1}^{+}\sqrt{-g}h_{\beta}^{0}\gamma^{\beta}\gamma^{0}\left(
O\psi_{2}\right)  =\int\left(  O\psi_{1}\right)  ^{+}\sqrt{-g}h_{\beta}%
^{0}\gamma^{\beta}\gamma^{0}\left(  \psi_{2}\right) \label{SP1}%
\end{equation}
It means that matricially
\begin{equation}
O^{+}U=UO\label{OU}%
\end{equation}
where "$+$" denotes from now, the usual Hermitic conjugation (transposition
and complex conjugation), that is, the hermitic conjugate with respect to the
scalar product in flat space denoted $\left\langle \mid\right\rangle $ and
defined by
\begin{equation}
\left\langle \psi_{1}(t)\mid\psi_{2}\left(  t\right)  \right\rangle =\int
\psi_{1}^{+}\psi_{2}d^{3}x\label{SP0}%
\end{equation}
Unfortunatly the definition Eq. $\left(  \ref{SP1}\right)  $ turns out to be
untractable for pratical computations. Actually, for the sake of the
diagonalization procedure, we aim at working with matrices which are Hermitian
with respect to the usual transpose and complex conjugate operation Eq.
$\left(  \ref{SP0}\right)  $, so that the diagonalization can be performed
through a unitary matrix in the usual sense. To do so, notice that if $O$ is
Hermitian for Eq. $\left(  \ref{SP}\right)  $, then Eq. $\left(
\ref{OU}\right)  $ implies that $U^{\frac{1}{2}}OU^{-\frac{1}{2}}$ is
Hermitian for Eq. $\left(  \ref{SP0}\right)  $. Thus, starting with the
Hamiltonian $\hat{H}$ defined in Eq. $\left(  \ref{H1}\right)  $, $U^{\frac
{1}{2}}\hat{H}U^{-\frac{1}{2}}$ is Hermitian in the usual sense and can be
diagonalized through a standard unitary matrix (that is unitary for $\left(
\ref{SP0}\right)  $).

The Hamiltonian of interest for us will thus be $U^{\frac{1}{2}}\hat
{H}U^{-\frac{1}{2}}$. It's hermiticity in the usual sense allows us to write:
\begin{align}
U^{\frac{1}{2}}\hat{H}U^{-\frac{1}{2}}  & =\frac{1}{2}U^{\frac{1}{2}}\hat
{H}U^{-\frac{1}{2}}+\frac{1}{2}\left(  U^{\frac{1}{2}}\hat{H}U^{-\frac{1}{2}%
}\right)  ^{+}\nonumber\\
& =\frac{1}{2}\left(  \hat{H}+\hat{H}^{+}\right)  +\frac{1}{2}\left[
U^{\frac{1}{2}},\hat{H}\right]  U^{-\frac{1}{2}}+\frac{1}{2}U^{-\frac{1}{2}%
}\left[  \hat{H},U^{\frac{1}{2}}\right] \label{dev}%
\end{align}
The non unitarity of the transformation $U$ is not problematic here, since it
is precisely used to change the metric from curved to flat scalar product, and
moreover $\hat{H}$ and $U^{\frac{1}{2}}\hat{H}U^{-\frac{1}{2}}$ have the same
spectrum. In the case of a static metric (time independent) and satisfying
$h_{\mu}^{0}=f\left(  \mathbf{R}\right)  \delta_{\mu}^{0}$, for a certain
position dependent function $f\left(  \mathbf{R}\right)  $, the transformation
$U^{\frac{1}{2}}$ reduces to the multiplication by a function of $\mathbf{R}$.
Then, using Eq. $\left(  \ref{H1}\right)  $ for $\hat{H}$, Eq. $\left(
\ref{dev}\right)  $ simplifies easily to:
\[
U^{\frac{1}{2}}\hat{H}U^{-\frac{1}{2}}=\frac{1}{2}U^{\frac{1}{2}}\hat
{H}U^{-\frac{1}{2}}+\frac{1}{2}\left(  U^{\frac{1}{2}}\hat{H}U^{-\frac{1}{2}%
}\right)  ^{+}=\frac{1}{2}\left(  \hat{H}+\hat{H}^{+}\right)
\]
It is this form that was considered in \cite{Photon}\cite{Pierregravit1}. If
in addition the metrics is diagonal, one recovers the transformation studied
in \cite{OBUKHOV}.

Independently of the practical advantages of the flat scalar product, there is
an other and deeper reason to transform the Hamiltonian to the flat space.
Actually, when diagonalizing the Hamiltonian with respect to Eq. $\left(
\ref{SP0}\right)  $\ the diagonal subspaces of up and down spinors will appear
to be obviously orthogonal. This is of course not the case for the scalar
product Eq. $\left(  \ref{SP}\right)  $ which mixes both subspaces. As a
consequence diagonalizing with respect to the curved space scalar product does
not lead to a clear separation between particles and antiparticles.

\subsection{Unitarity}

We will end up this section by stressing the fact that for a non-static metric
the Hamiltonian and the time evolution operator do not coincide in the flat
representation, and in addition, the time evolution operator cannot be made
Hermitian. To make this point clearer, consider the Schr\"{o}dinger equation
\[
i\hbar\frac{\partial}{\partial t}\Psi=H\Psi
\]
Applying the transformation $U^{\frac{1}{2}}$ yields,%
\[
\frac{\partial}{\partial t}\Psi^{\prime}=\left(  U^{\frac{1}{2}}\left(
t\right)  HU^{-\frac{1}{2}}\left(  t\right)  -i\hbar U^{\frac{1}{2}}\left(
t\right)  \frac{\partial}{\partial t}U^{-\frac{1}{2}}\left(  t\right)
\right)  \Psi^{\prime}
\]
with $\Psi^{\prime}=U^{\frac{1}{2}}\Psi$. As a consequence the evolution for
$\Psi^{\prime}$, is given by the operator
\begin{equation}
H^{e}=U^{\frac{1}{2}}\left(  t\right)  HU^{-\frac{1}{2}}\left(  t\right)
-i\hbar U^{\frac{1}{2}}\left(  t\right)  \frac{\partial}{\partial t}%
U^{-\frac{1}{2}}\left(  t\right) \label{ev}%
\end{equation}
One would like $H^{e}$ to be hermitian for the scalar product Eq. $\left(
\ref{SP0}\right)  $. However, due to the non unitarity of $U^{\frac{1}{2}}$,
the contribution $i\hbar U^{\frac{1}{2}}\left(  t\right)  \frac{\partial
}{\partial t}U^{-\frac{1}{2}}\left(  t\right)  $ is not. The reason for this
non unitarity tracks back to the dependence in $t$ of the scalar product Eq.
$\left(  \ref{SP}\right)  $, so that the norm of a wave function is not
preserved in time. Actually starting with an initial condition $\Psi^{\prime
}\left(  t_{0}\right)  $, the solution for the Schroedinger equation is
\[
\Psi^{\prime}\left(  t_{1}\right)  =T\exp\left(  \int_{t_{0}}^{t_{1}}%
H^{e}\left(  t\right)  dt\right)  \Psi^{\prime}\left(  t_{0}\right)
=U^{\frac{1}{2}}\left(  t_{1}\right)  T\exp\left(  \int_{t_{0}}^{t_{1}%
}H\left(  t\right)  dt\right)  U^{-\frac{1}{2}}\left(  t_{0}\right)
\Psi^{\prime}\left(  t_{0}\right)
\]
where $T$ is the time ordered exponential. Assuming the norm $\Psi^{\prime
}\left(  t_{0}\right)  $ to be equal to $1$, $\Psi^{\prime}\left(
t_{1}\right)  $ is easily seen to have a norm (respectively to Eq. $\left(
\ref{SP}\right)  $) different from one. That can be checked easily on an
infinitesimal timeslice, $t_{1}=t_{0}+\Delta t$. Indeed
\begin{align}
\left\langle \Psi^{\prime}\left(  t_{1}\right)  \right\vert \left\vert
\Psi^{\prime}\left(  t_{1}\right)  \right\rangle  & =\left\langle U^{\frac
{1}{2}}\left(  t_{1}\right)  \exp\left(  -iH\left(  t_{0}\right)  \Delta
t\right)  \Psi\left(  t_{0}\right)  \mid U^{\frac{1}{2}}\left(  t_{1}\right)
\exp\left(  iH\left(  t_{0}\right)  \Delta t\right)  \Psi\left(  t_{0}\right)
\right\rangle \nonumber\\
& =\left\langle \exp\left(  -iH\left(  t_{0}\right)  \Delta t\right)
\Psi\left(  t_{0}\right)  \mid U\left(  t_{1}\right)  \exp\left(  iH\left(
t_{0}\right)  \Delta t\right)  \Psi\left(  t_{0}\right)  \right\rangle
\nonumber\\
& =\left\langle \exp\left(  -iH\left(  t_{0}\right)  \Delta t\right)
\Psi\left(  t_{0}\right)  \mid\exp\left(  iH\left(  t_{0}\right)  \Delta
t\right)  \Psi\left(  t_{0}\right)  \right\rangle _{U\left(  t_{1}\right)
}\label{scal}%
\end{align}
and this is different from $1$ since,
\begin{align}
& \left\langle \exp\left(  -iH\left(  t_{0}\right)  \Delta t\right)
\Psi\left(  t_{0}\right)  \mid\exp\left(  iH\left(  t_{0}\right)  \Delta
t\right)  \Psi\left(  t_{0}\right)  \right\rangle _{U\left(  t_{1}\right)
}\nonumber\\
& =\left\langle \exp\left(  -iH\left(  t_{0}\right)  \Delta t\right)
\Psi\left(  t_{0}\right)  \mid U\left(  t_{0}\right)  \exp\left(  iH\left(
t_{0}\right)  \Delta t\right)  \Psi\left(  t_{0}\right)  \right\rangle
+\left\langle \Psi\left(  t_{0}\right)  \mid\frac{\partial}{\partial
t}U\left(  t_{0}\right)  \Delta t\Psi\left(  t_{0}\right)  \right\rangle \\
& =\left\langle \exp\left(  -iH\left(  t_{0}\right)  \Delta t\right)
\Psi\left(  t_{0}\right)  \mid\exp\left(  iH\left(  t_{0}\right)  \Delta
t\right)  \Psi\left(  t_{0}\right)  \right\rangle _{U\left(  t_{0}\right)
}+\left\langle \Psi\left(  t_{0}\right)  \mid\frac{\partial}{\partial
t}U\left(  t_{0}\right)  \Delta t\Psi\left(  t_{0}\right)  \right\rangle
\label{dyna}%
\end{align}
the first term is equal to $1$, actually $\Psi\left(  t_{0}\right)  $ is of
norm $1$ for $\left\langle \mid\right\rangle _{U\left(  t_{0}\right)  }$ and
$\exp\left(  iH\left(  t_{0}\right)  \Delta t\right)  $ is unitary for this
scalar product. As a consequence, $\left\langle \Psi^{\prime}\left(
t_{1}\right)  \right\vert \left\vert \Psi^{\prime}\left(  t_{1}\right)
\right\rangle $ differs from one and $H^{e}$ is non unitary. The reason is
clear from Eq. $\left(  \ref{scal}\right)  $: a vector of norm $1$ for
$\left\langle \mid\right\rangle _{U\left(  t_{0}\right)  }$ is transported to
a vector, that has no more norm $1$ for $\left\langle \mid\right\rangle
_{U\left(  t_{1}\right)  }$. During the evolution, the matrix $U$ defining the
scalar product has changed too, and the non hermitian connexion term $-i\hbar
U^{\frac{1}{2}}\left(  t\right)  \frac{\partial}{\partial t}U^{-\frac{1}{2}%
}\left(  t\right)  $ tracks the change of metric between $t$ and $t+\Delta t$.

Therefore, the time evolution of the state is non-unitary. In the rest of the
paper we focus on the diagonalization of the energy operator $U^{\frac{1}{2}%
}\left(  t\right)  HU^{-\frac{1}{2}}\left(  t\right)  $ this one being
Hermitian, although the diagonalization of $-i\hbar U^{\frac{1}{2}}\left(
t\right)  \frac{\partial}{\partial t}U^{-\frac{1}{2}}\left(  t\right)  $ is
provided for the sake of completness in appendix B.

\subsection{Transformation to the flat space}

We thus now focus on the Hamiltonian Eq. $\left(  \ref{H1}\right)  $, and
compute explicitly $U^{\frac{1}{2}}\hat{H}U^{-\frac{1}{2}}$ which as shown is
hermitian in the usual sense. The transformation $U$ is given by
\[
U=\sqrt{-g}\left(  h_{0}^{0}+h_{\beta}^{0}\alpha^{\beta}\right)  =\sqrt
{-g}h_{0}^{0}\left(  1+\alpha^{\beta}h_{\beta}^{0}/h_{0}^{0}\right)
\]
One can then deduce
\[
U^{\frac{1}{2}}=f\left(  1+u_{\beta}\alpha^{\beta}\right)
\]
and thus
\[
U^{-\frac{1}{2}}=\frac{1}{f\left(  1-u^{2}\right)  }\left(  1-u_{\beta}%
\alpha^{\beta}\right)
\]
where $u^{2}=u_{\beta}u^{\beta}$ with
\[
u_{\beta}=\frac{h_{\beta}^{0}/h_{0}^{0}}{\sqrt{-g}h_{0}^{0}\left(
1+\sqrt{1-\frac{h_{\beta}^{0}h^{0\beta}}{\left(  h_{0}^{0}\right)  ^{2}}%
}\right)  }.
\]
and%
\[
f=\left(  \sqrt{-g}h_{0}^{0}\frac{1+\sqrt{1-\frac{h_{\beta}^{0}h^{0\beta}%
}{\left(  h_{0}^{0}\right)  ^{2}}}}{2}\right)  ^{1/2}
\]
The greek (lorentzian) indices are assumed now to run only from $1$ to $3$. As
a consequence the Hamiltonian $H=U^{\frac{1}{2}}\hat{H}U^{-\frac{1}{2}}$ given
by Eq. $\left(  \ref{dev}\right)  $ reads
\[
H=\frac{1}{2}\left(  H+H^{+}\right)  +\frac{1}{2}f\left(  1+u_{\beta}%
\alpha^{\beta}\right)  \left[  H,\frac{1}{f\left(  1-u^{2}\right)  }\left(
1-u_{\beta}\alpha^{\beta}\right)  \right]  -\frac{1}{2}\left[  H^{+},\left(
1-u_{\beta}\alpha^{\beta}\right)  \frac{1}{f\left(  1-u^{2}\right)  }\right]
\left(  1+u_{\beta}\alpha^{\beta}\right)  f
\]
which can be rewritten as
\begin{align}
H  & =\frac{1}{2}\left(  H+H^{+}\right)  -\frac{1}{2}\frac{\left(  1+u_{\beta
}\alpha^{\beta}\right)  }{\left(  1-u^{2}\right)  }\left[  H,u_{\beta}%
\alpha^{\beta}\right]  +\frac{1}{2}\left[  H^{+},u_{\beta}\alpha^{\beta
}\right]  \frac{\left(  1+u_{\beta}\alpha^{\beta}\right)  }{\left(
1-u^{2}\right)  }\nonumber\\
& -i\frac{\hbar}{2}\left(  1+u_{\beta}\alpha^{\beta}\right)  \left(
\mathbf{\nabla}_{\mathbf{P}}H\right)  .\mathbf{\nabla}_{\mathbf{R}}\left(
\frac{1}{f\left(  1-u^{2}\right)  }\right)  \left(  1-u_{\beta}\alpha^{\beta
}\right) \nonumber\\
& +i\frac{\hbar}{2}\left(  1-u_{\beta}\alpha^{\beta}\right)  \left(
\mathbf{\nabla}_{\mathbf{P}}H^{+}\right)  .\mathbf{\nabla}_{\mathbf{R}}\left(
\frac{1}{f\left(  1-u^{2}\right)  }\right)  \left(  1+u_{\beta}\alpha^{\beta
}\right) \nonumber
\end{align}
where $\mathbf{P}$ and $\mathbf{R}$ are the canonical momentum and position
operator satisfying $\left[  R^{i},P_{j}\right]  =i\hbar\delta_{j}^{i}.$ The
Hamiltonian can also be written as%

\begin{align}
H  & =\frac{1}{2}\left(  \frac{1}{\left(  1-u^{2}\right)  }H+H^{+}\frac
{1}{\left(  1-u^{2}\right)  }\right)  -\frac{1}{2}\frac{1}{\left(
1-u^{2}\right)  }\left[  H,u_{\beta}\alpha^{\beta}\right]  +\frac{1}{2}\left[
H^{+},u_{\beta}\alpha^{\beta}\right]  \frac{1}{\left(  1-u^{2}\right)
}\nonumber\\
& -\frac{1}{2}u_{\beta}\alpha^{\beta}\left(  \frac{1}{\left(  1-u^{2}\right)
}H+H^{+}\frac{1}{\left(  1-u^{2}\right)  }\right)  u_{\beta}\alpha^{\beta
}-i\frac{\hbar}{2}\left[  u_{\beta}\alpha^{\beta},\left(  \mathbf{\nabla
}_{\mathbf{P}}H\right)  .\mathbf{\nabla}_{\mathbf{R}}\left(  \frac{1}{f\left(
1-u^{2}\right)  }\right)  \right] \label{HHApp}%
\end{align}
At this level, the computation of the last expression turns out to be quite
technical is fully developped in Appendix A. The result is given by the
following expression
\begin{align}
H  & =\frac{1}{2}\alpha^{\beta}\tilde{H}_{\beta}^{i}\left(  P_{i}%
+\hbar\varepsilon_{\varrho\beta\gamma}\frac{\tilde{\Gamma}_{i}^{\varrho\beta}%
}{4}\Sigma^{\gamma}\right)  +\frac{1}{2}\left(  P_{i}+\hbar\varepsilon
_{\varrho\beta\gamma}\frac{\tilde{\Gamma}_{i}^{\varrho\beta}}{4}\Sigma
^{\gamma}\right)  \tilde{H}_{\beta}^{i}\alpha^{\beta}\nonumber\\
& +\beta\tilde{m}+\frac{1}{2}g^{i}P_{i}+P_{i}g^{i}+\hbar\left(  \mathbf{\tilde
{\Gamma}}^{0}+\mathbf{\Gamma}^{e}\right)  .\mathbf{\Sigma}+\hbar\frac{\left(
g_{00}\left(  \mathbf{h}^{0}\times\mathbf{h}^{i}\right)  -\mathbf{u\times
H}^{i}\right)  }{\left(  1-u^{2}\right)  }.\left(  \nabla_{i}\mathbf{u}%
\right)  J\label{HH}%
\end{align}
where from now on, all indices $i$, $\beta$,$\rho...$ are only spatial and run
from $1$ to $3$, but roman indices are raised and lowered by the meric
$g_{ij}$ and greek indices by the lorentzian metrics $\eta_{\alpha\beta}.$ The
several notations introduced above are given by the following expressions%
\begin{align*}
\tilde{H}_{\beta}^{i}  & =H_{\beta}^{i}+\frac{2g_{00}\left(  h_{\delta}%
^{0}h_{\beta}^{i}-h_{\beta}^{0}h_{\delta}^{i}\right)  u_{\delta}}{\left(
1-u^{2}\right)  }\\
\hat{\Gamma}_{i}^{\varrho\beta}  & =\Gamma_{i}^{\varrho\beta}-\frac{1}%
{4}\varepsilon_{\text{ \ \ }\gamma}^{\varrho\beta}\left(  H^{-1}\right)
_{i}^{\kappa}g_{00}h_{\delta}^{0}h_{\eta}^{i}\varepsilon_{\text{ \ \ \ }%
\kappa}^{\delta\eta}\Gamma_{j}^{0\gamma}\\
\tilde{\Gamma}_{i}^{\varrho\beta}  & =\left(  \tilde{H}^{-1}\right)
_{i}^{\eta}H_{\eta}^{j}\hat{\Gamma}_{j}^{\varrho\beta}\\
\hat{\Gamma}_{\gamma}^{0}  & =\frac{1}{4}\left(  \varepsilon_{\varrho
\beta\gamma}\Gamma_{0}^{\varrho\beta}+g_{00}g^{0i}\varepsilon_{\varrho
\beta\gamma}\Gamma_{i}^{\varrho\beta}+\varepsilon_{\text{ }\nu\gamma}^{\beta
}H_{\beta}^{i}\Gamma_{i}^{0\nu}\right) \\
\tilde{\Gamma}_{\gamma}^{0}  & =\frac{\left(  1+u^{2}\right)  \delta_{\gamma
}^{\eta}-u^{\eta}u_{\gamma}}{\left(  1-u^{2}\right)  }\hat{\Gamma}_{\eta}%
^{0}+\frac{1}{2}\left(  \left(  2H_{\delta}^{i}\mathbf{\hat{\Gamma}}%
_{i}^{\delta}\mathbf{+\Gamma}_{0}^{0}\mathbf{-}g_{00}g^{0i}\mathbf{\Gamma}%
_{i}^{0}\right)  \times\mathbf{u}\right)  _{\gamma}\\
\mathbf{\Gamma}_{\gamma}^{e}  & =-\left(  \mathbf{u\times H}^{i}\right)
_{\gamma}\left(  \frac{\nabla_{R_{i}}\left(  f\left(  1-u^{2}\right)  \right)
}{f^{2}\left(  1-u^{2}\right)  ^{3}}\right)  +\frac{\left(  \mathbf{u\times
}\nabla_{i}\left(  \mathbf{H}^{i}\right)  \right)  _{\gamma}}{\left(
1-u^{2}\right)  }-\frac{\left(  1+u^{2}\right)  \delta_{\gamma}^{\eta}%
-u^{\eta}u_{\gamma}}{\left(  1-u^{2}\right)  }\nabla_{R_{i}}\left(
g_{00}\left(  \mathbf{h}^{0}\times\mathbf{h}^{i}\right)  _{\eta}\right)
\end{align*}
where we used vectorial notations $\left(  \mathbf{h}^{0}\right)  _{\alpha
}=h_{\alpha}^{0}$, $\left(  \mathbf{h}^{i}\right)  _{\alpha}=h_{\alpha}^{i}$,
as well as the vectors $\left(  \mathbf{H}^{i}\right)  _{\beta}\equiv
H_{\beta}^{i}=g_{00}h_{0}^{0}\left(  h_{\beta}^{i}-\frac{h_{\beta}^{0}%
h_{0}^{i}}{h_{0}^{0}}\right)  ,$ the following various vectors $\left(
\mathbf{\tilde{\Gamma}}_{i}^{\delta}\right)  ^{\beta}=\tilde{\Gamma}%
_{i}^{\delta\beta},$ $\left(  \hat{\Gamma}_{i}^{\delta}\right)  ^{\beta}%
=\hat{\Gamma}_{i}^{\delta\beta}$, $\left(  \mathbf{\Gamma}_{i}^{0}\right)
^{\beta}=\Gamma_{i}^{0\beta},$ and $\left(  \mathbf{\Gamma}_{0}^{0}\right)
^{\beta}=\Gamma_{0}^{0\beta}$. The matrix $J$ is given by $J=\left(
\begin{array}
[c]{cc}%
0 & I_{2\times2}\\
I_{2\times2} & 0
\end{array}
\right)  $ and the effective mass by $\tilde{m}=mg_{00}h_{0}^{0}\left(
\frac{1+u^{2}}{1-u^{2}}\right)  $ as well as $g^{i}=g_{00}g^{0i}.$ The
expression Eq.$\left(  \ref{HH}\right)  $ for the energy operator $H$ will be
the one to diagonalize in the next section.

\section{Semi-Classical Energy}

The semiclassical diagonalization the Hamiltonian Eq. $\left(  \ref{HH}%
\right)  $ is expected to lead to an effective Hamiltonian with gauge fields
resulting from the back reaction of the spin degree of freedom (fast) on the
translational momentum which can be treated semiclassically for slowly varying
enough inhomogeneities. Indeed, the emergence of gauge fields is a general
feature of systems providing fast and slow degrees of freedom. The purpose of
ref \cite{BO}, was to investigate the origin of quantum gauge fields and
forces by considering the diagonalization of an arbitrary matrix valued
quantum Hamiltonian. To be precise, by diagonalization we mean the derivation
of an effective in-band Hamiltonian made of block-diagonal energy subspaces.
This approach, based on a new differential calculus on a non-commutative space
where $\hbar$ plays the role of running parameter, leads to an in-band energy
operator that can be obtained systematically up to arbitrary order in $\hbar
$\textbf{. }Particularly important for our purpose, it has been possible, for
an arbitrary Hamiltonian $H(\mathbf{R},\mathbf{P})$ with the canonical
coordinates and momentum $\left[  R_{i},P^{j}\right]  =i\hbar\delta_{i}^{j},$
to obtain the corresponding diagonal representation $\varepsilon\left(
\mathbf{r,p}\right)  $ to order $\hbar^{2} $, in terms of non-canonical
coordinates and momentum ($\mathbf{r,p}$\textbf{)} defined later and
commutators between gauge fields. The method is quite involved, so that in the
present paper we restrict ourself to the semiclassical approximation (order
$\hbar$).

The mathematical difficulty in performing the diagonalization of $H$ comes
from the intricate entanglement of noncommuting operators due to the canonical
relation $\left[  R_{i},P^{j}\right]  =i\hbar\delta_{i}^{j}.$ In
\cite{PIERRE1} starting with a very general but time independent
$H(\mathbf{R,P})$ and by considering $\hbar$ as a running parameter, we
related the in-band Hamiltonian $VHV^{+}=\varepsilon\left(  \mathbf{X}\right)
$ and the unitary transforming matrix $V\left(  \mathbf{X}\right)  $ (where
$\mathbf{X}\equiv(\mathbf{R,P})$) to their classical expressions through
integro-differential operators, i.e. $\varepsilon\left(  \mathbf{X}\right)
=\widehat{O}\left(  \varepsilon(\mathbf{X}_{0})\right)  $ and $V\left(
\mathbf{X}\right)  =\widehat{N}\left(  V(\mathbf{X}_{0})\right)  $, where in
the matrices $\varepsilon(\mathbf{X}_{0})$ and $V(\mathbf{X}_{0}),$ the
dynamical operators $\mathbf{X}$ are replaced by classical commuting variables
$\mathbf{X}_{0}=(\mathbf{R}_{0},\mathbf{P}_{0}).$

The only requirement of the method is therefore the knowledge of
$V(\mathbf{X}_{0})$ which gives the diagonal form $\varepsilon(\mathbf{X}%
_{0}).$ Generally, these equations do not allow to find directly
$\varepsilon\left(  \mathbf{X}\right)  $, $V\left(  \mathbf{X}\right)  $,
however, they allow us to produce the solution recursively in a series
expansion in $\hbar.$ With this assumption that both $\varepsilon$ and $V$ can
be expanded in power series of $\hbar,$ we determined in \cite{PIERRE1}, the
explicit $n$-th in band energy to order $\hbar$ for an arbitrary given
Hamiltonian%
\begin{equation}
\varepsilon_{n}\left(  \mathbf{r,p}\right)  =\varepsilon_{0,n}\left(
\mathbf{r,p}\right)  +\frac{i\hbar}{2}\mathcal{P}_{n}\left\{  \left[
\varepsilon_{0}\left(  \mathbf{r,p}\right)  ,\mathcal{A}^{R_{l}}\right]
\mathcal{A}^{P_{l}}-\left[  \varepsilon_{0}\left(  \mathbf{r,p}\right)
,\mathcal{A}^{P_{l}}\right]  \mathcal{A}^{R_{l}}\right\}  +O(\hbar
^{2})\label{EN}%
\end{equation}
where $\mathcal{A}_{\mathbf{R}}=i\hbar V\nabla_{\mathbf{P}}V^{+}$ and
$\mathcal{A}_{\mathbf{P}}=-i\hbar V\nabla_{\mathbf{R}}V^{+}$ with $V$ the
diagonalizing matrix. The operator $\mathcal{P}_{n}$ has the meaning of the
projection on the $n$-th energy subspace. The new non-canonical dynamical
operators $\mathbf{r}$ and $\mathbf{p}$ depend on gauge fields $\mathbf{A}%
_{R}=\mathcal{P}_{n}(\mathcal{A}_{\mathbf{R}})$ and $\mathbf{A}_{P}%
$=$\mathcal{P}_{n}(\mathcal{A}_{\mathbf{P}})$ similarly to electromagnetism,
as we have $\mathbf{r}=i\hbar\partial_{\mathbf{p}}+\mathbf{A}_{R}$ and
$\mathbf{p}=\mathbf{P}+\mathbf{A}_{P}$. These gauge invariant quantities\ not
only are emerging naturally but are also necessary to have a gauge invariant
energy Eq. $\left(  \ref{EN}\right)  $. The operator $\varepsilon_{0}\left(
\mathbf{r,p}\right)  $ is the diagonal energy obtained at zero order
($\hbar^{0})$ in which the classical variable $\mathbf{R}_{0}$ and
$\mathbf{P}_{0}$ are replaced by new non-commuting operators $\mathbf{r,p}$.
Therfore the first step consists in finding the matrix $V(\mathbf{X}_{0})$
which diagonalizes the "classical" Hamiltonian $H(\mathbf{X}_{0}).$

\subsection{Zero Order Diagonalization.}

For practical purpose we introduce the three dimensional effective metric
$G^{ij}=\tilde{H}_{\alpha}^{i}\tilde{H}_{\beta}^{j}\delta^{\alpha\beta}$ as
well as the gravity coupled momentum $\tilde{P}_{\alpha}\mathbf{=}\tilde
{H}_{\alpha}^{i}(P_{i}+\frac{\hbar}{4}\varepsilon_{\varrho\beta\gamma}%
\tilde{\Gamma}_{i}^{\varrho\beta}\sigma^{\gamma}).$\ 

As shown in \cite{Pierregravit1}, the classical block-diagonalization of the
Hamiltonian Eq. $\left(  \ref{HH}\right)  $, but without the last term
(corresponding to the static case), can be performed by the following unitary
FW-like matrix (denoted $F_{0}$ for Foldy-Wouthuysen)
\[
F_{0}(\mathbf{\tilde{P})}=D\left(  E_{0}+\tilde{m}+c\frac{1}{2}\beta\left(
\mathbf{\alpha}.\mathbf{\tilde{P}+\tilde{P}}^{+}\mathbf{.\alpha}\right)
+N\right)  \mathbf{/}\sqrt{2E_{0}\left(  E_{0}+\tilde{m}\right)  }
\]
with $E_{0}=\sqrt{\left(  \frac{\mathbf{\alpha}.\mathbf{\tilde{P}+\tilde{P}%
}^{+}\mathbf{.\alpha}}{2}\right)  ^{2}+\tilde{m}^{2}}$, $N=\frac{\hbar}%
{4}\frac{i\mathbf{\alpha}.\left(  \mathbf{\tilde{P}\times}\left(
\mathbf{\tilde{\Gamma}}^{0}+\hbar\mathbf{\Gamma}^{e}\right)  \right)  }{E_{0}%
}$, and $D=1+\ \frac{\hbar}{4}\beta\frac{\left(  \mathbf{\tilde{P}\times
}\left(  \mathbf{\tilde{\Gamma}}^{0}+\hbar\mathbf{\Gamma}^{e}\right)  \right)
\times\mathbf{\tilde{P}}}{2E_{0}^{2}(E_{0}+\tilde{m})}$.

Indeed one can easily check that
\begin{align}
F_{0}HF_{0}{}^{+}  & =\beta\sqrt{P_{i}G^{ij}P_{j}+\hbar\varepsilon
_{\alpha\beta\gamma}\tilde{\Gamma}_{j}^{\alpha\beta}\Sigma^{\gamma}G^{ij}%
P_{i}+\tilde{m}^{2}}\nonumber\\
& +\frac{\hbar}{4E_{0}}\left(  \mathbf{\tilde{\Gamma}}^{0}+\mathbf{\Gamma}%
^{e}\right)  .\left(  \tilde{m}\mathbf{\Sigma}+\frac{\left(  \mathbf{\Sigma
}.\mathbf{\tilde{H}}^{i}P_{i}\right)  \mathbf{\tilde{H}}^{i}P_{i}}%
{(E_{0}+\tilde{m})}\right)  +\frac{1}{2}g^{i}P_{i}+\frac{1}{2}P_{i}%
g^{i}\nonumber\\
& +F_{0}\left(  \frac{\left(  g_{00}\left(  \mathbf{h}^{0}\times\mathbf{h}%
^{i}\right)  -\mathbf{u\times H}^{i}\right)  }{\left(  1-u^{2}\right)
}.\left(  \nabla_{i}\mathbf{u}\right)  J\right)  F_{0}{}^{+}\label{EO}%
\end{align}
with $\tilde{\Gamma}_{i\gamma}=\varepsilon_{\varrho\beta\gamma}\tilde{\Gamma
}_{i}^{\varrho\beta}$. All contributions are block-diagonal except the last
term which was not present before in the case of a static metrics
\cite{Pierregravit1}. The proof of this block diagonalization relies on the
simple fact that for classical variables $\mathbf{X}_{0},$ the matrices
$\tilde{H}_{\alpha}^{i}$ and $\tilde{\Gamma}_{i}^{\alpha\beta}$ are
independent of both the momentum and position, $\beta$ and $\mathbf{\alpha
.\tilde{P}}$ anticommute and in the Taylor expansion of $E_{0}$ all terms
commute with $\beta$ and $\mathbf{\alpha.\tilde{P}+\mathbf{\tilde{P}}%
^{+}\mathbf{.\alpha}}$.

As said before, the last term in Eq.$\left(  \ref{EO}\right)  $ is non
diagonal and must treated specifically. Actuallt, one can apply a second
unitary transformation that will cancel the non diagonal contributions of
$F_{0}(\frac{\left(  g_{00}\left(  \mathbf{h}^{0}\times\mathbf{h}^{i}\right)
-\mathbf{u\times H}^{i}\right)  }{\left(  1-\mathbf{u}^{2}\right)  }.\left(
\nabla_{i}\mathbf{u}\right)  J)F_{0}{}^{+}$ without affecting the rest of the
diagonalized Hamiltonian to the first order in $\hbar$. The explicit form of
this transformation is :%
\[
F_{0}^{\prime}=1-\mathcal{P}_{-}\left(  F_{0}\frac{\left(  g_{00}\left(
\mathbf{h}^{0}\times\mathbf{h}^{i}\right)  -\mathbf{u\times H}^{i}\right)
}{\left(  1-\mathbf{u}^{2}\right)  }.\left(  \nabla_{i}\mathbf{u}\right)
JF_{0}{}^{+}\right)  \frac{\beta}{2\sqrt{P_{i}G^{ij}P_{j}+\tilde{m}^{2}}}
\]
$\mathcal{P}_{-}$ being the projection outside the diagonal. One can check
that $F_{0}^{\prime}-1$ is antihermitian so that $F_{0}^{\prime}$ is unitary
to the first order. As a consequence, the composition of the two unitary
transformations yields the following diagonal energy operator :%
\begin{align}
\varepsilon_{0}\left(  \mathbf{R},\mathbf{P,}t\right)   & =F_{0}^{\prime}%
F_{0}\hat{H}_{0}F_{0}{}^{+}F_{0}^{\prime+}=\beta\sqrt{P_{i}G^{ij}P_{j}%
+\hbar\varepsilon_{\alpha\beta\gamma}\tilde{\Gamma}_{j}^{\alpha\beta}%
\Sigma^{\gamma}G^{ij}P_{i}+\tilde{m}^{2}}\nonumber\\
& +\frac{\hbar}{4E_{0}}\left(  \mathbf{\tilde{\Gamma}}^{0}+\mathbf{\Gamma}%
^{e}\right)  .\left(  \tilde{m}\mathbf{\Sigma}+\frac{\left(  \mathbf{\Sigma
}.\mathbf{\tilde{H}}^{i}P_{i}\right)  \mathbf{\tilde{H}}^{i}P_{i}}%
{(E_{0}+\tilde{m})}\right)  +\frac{1}{2}g^{i}P_{i}+\frac{1}{2}P_{i}%
g^{i}\nonumber\\
& +\mathcal{P}_{+}\left(  F_{0}\left(  \frac{\left(  g_{00}\left(
\mathbf{h}^{0}\times\mathbf{h}^{i}\right)  -\mathbf{u\times H}^{i}\right)
}{\left(  1-\mathbf{u}^{2}\right)  }.\left(  \nabla_{i}\mathbf{u}\right)
J\right)  F_{0}{}^{+}\right)
\end{align}
The last term is explicitely given by:%
\begin{align*}
& \mathcal{P}_{+}\left(  F_{0}\left(  \frac{\left(  g_{00}\left(
\mathbf{h}^{0}\times\mathbf{h}^{i}\right)  -\mathbf{u\times H}^{i}\right)
}{\left(  1-\mathbf{u}^{2}\right)  }.\left(  \nabla_{i}\mathbf{u}\right)
J\right)  F_{0}{}^{+}\right) \\
& =D\mathcal{P}_{+}\left(  E_{0}+\tilde{m}+c\beta\left(  \mathbf{\alpha
}.\mathbf{\tilde{P}}\right)  \right)  \left(  \left(  E_{0}+V(\mathbf{r}%
)m\right)  J+c\beta\left(  \mathbf{\Sigma}.\mathbf{\tilde{P}}\right)  \right)
\frac{\hbar\left(  g_{00}\left(  \mathbf{h}^{0}\times\mathbf{h}^{i}\right)
-\mathbf{u\times H}^{i}\right)  .\left(  \nabla_{i}\mathbf{u}\right)  }%
{2E_{0}\left(  E_{0}+\tilde{m}\right)  }D^{+}\\
& =\hbar c\beta\frac{\left(  g_{00}\left(  \mathbf{h}^{0}\times\mathbf{h}%
^{i}\right)  -\mathbf{u\times H}^{i}\right)  .\left(  \nabla_{i}%
\mathbf{u}\right)  }{2E_{0}\left(  1-\mathbf{u}^{2}\right)  }\mathbf{\tilde
{P}}.\mathbf{\Sigma=}\hbar c\beta\frac{\left(  \left(  f-1\right)
g_{00}\left(  \mathbf{h}^{0}\times\mathbf{h}^{i}\right)  \right)  .\left(
\nabla_{i}\mathbf{u}\right)  }{2fE_{0}\left(  1-\mathbf{u}^{2}\right)
}\mathbf{\tilde{P}}.\mathbf{\Sigma}%
\end{align*}
Ultimately, the diagonalization process yields:%
\begin{align}
\varepsilon_{0}\left(  \mathbf{R},\mathbf{P,}t\right)   & =F_{0}^{\prime}%
F_{0}\hat{H}_{0}F_{0}{}^{+}F_{0}^{\prime+}\nonumber\\
& =\beta\sqrt{P_{i}G^{ij}P_{j}+\hbar\varepsilon_{\alpha\beta\gamma}%
\tilde{\Gamma}_{j}^{\alpha\beta}\Sigma^{\gamma}G^{ij}P_{i}+\tilde{m}^{2}%
}\nonumber\\
& +\hbar\left(  \mathbf{\tilde{\Gamma}}^{0}+\mathbf{\Gamma}^{e}\right)
.\left(  \tilde{m}\mathbf{\Sigma}+\frac{\left(  \mathbf{\Sigma}.\mathbf{\tilde
{H}}^{i}P_{i}\right)  \mathbf{\tilde{H}}^{i}P_{i}}{(E_{0}+\tilde{m})}\right)
+\frac{1}{2}g^{i}P_{i}+\frac{1}{2}P_{i}g^{i}+\hbar c\beta\frac{\left(  \left(
f-1\right)  g_{00}\left(  \mathbf{h}^{0}\times\mathbf{h}^{i}\right)  \right)
.\left(  \nabla_{i}\mathbf{u}\right)  }{2fE_{0}\left(  1-\mathbf{u}%
^{2}\right)  }\mathbf{\tilde{P}}.\mathbf{\Sigma}\label{ENclassic}%
\end{align}
where for the moment $\mathbf{R}$ and $\mathbf{P}$ are treated as classical
commuting quantities.

\subsection{First order in $\hbar$ diagonalization}

From expression Eq. $\left(  \ref{ENclassic}\right)  $ we can deduce the
diagonal energy operator for, let say, the particule subspace $\varepsilon
_{0,+}$. The semiclassical energy is given by Eq.\ $\left(  \ref{EN}\right)
$, where $\varepsilon_{0,+}$ corresponds to the positive energy subspace Eq.
$\left(  \ref{ENclassic}\right)  $ in which the classical variables
$\mathbf{R},\mathbf{P}$ are replaced by the quantum covariant ones
$\mathbf{r}=i\hbar\partial_{\mathbf{p}}+\hbar\mathbf{A}_{R}$ and
$\mathbf{p}=\mathbf{P}+\hbar\mathbf{A}_{P}.$ The explicit computation for the
Berry connections $\mathbf{A}_{R}=\mathcal{P}_{+}(\mathcal{A}_{\mathbf{R}})$
and $\mathbf{A}_{P}=\mathcal{P}_{+}(\mathcal{A}_{\mathbf{P}})$ with
$\mathcal{A}_{\mathbf{R}}=\mathcal{P}_{+}(i\hbar\left(  F_{0}^{\prime}%
F_{0}\right)  \nabla_{\mathbf{P}}(F_{0}^{\prime}F_{0})^{+})$ and
$\mathcal{A}_{\mathbf{P}}=\mathcal{P}_{+}(i\hbar\left(  F_{0}^{\prime}%
F_{0}\right)  \nabla_{\mathbf{R}}(F_{0}^{\prime}F_{0})^{+})$ gives the
components
\begin{equation}
A_{R_{k}}=c^{2}\frac{\varepsilon^{\alpha\beta\gamma}\tilde{H}_{\gamma}%
^{i}\tilde{P}_{\alpha}\mathbf{\Sigma}_{\beta}}{2E\left(  E+\tilde{m}\right)
}g_{ik}+o\left(  \hbar\right) \label{rGB}%
\end{equation}

\begin{equation}
A_{P_{k}}=-c^{2}\frac{\varepsilon^{\alpha\beta\gamma}\tilde{P}_{\alpha
}\mathbf{\Sigma}_{\beta}(\mathbf{\nabla}_{R_{k}}\tilde{P}_{\gamma})}{2E\left(
E+\tilde{m}\right)  }+o\left(  \hbar\right) \label{pGB}%
\end{equation}
where $E$ is the same as $E_{0}$ above, but now $\mathbf{R}$ is an operator
and $\tilde{H}_{\gamma}^{k}$ is the inverse matrix of $\tilde{H}_{\beta}^{i}$.

We also denote, for the rest of the paper, $\mathbf{\tilde{p}}$ to be the same
expression as $\mathbf{\tilde{P}}$ in which $\mathbf{R}$ and $\mathbf{P} $
have been replaced by $\mathbf{r}$ and $\mathbf{p}$, namely:
\[
\tilde{p}_{\alpha}\mathbf{=}\tilde{H}_{\alpha}^{i}(p_{i}+\frac{\hbar}%
{4}\varepsilon_{\varrho\beta\gamma}\tilde{\Gamma}_{i}^{\varrho\beta}%
\sigma^{\gamma})
\]
To complete the diagonalization we need to evaluate the quantity
\[
M_{+}=\frac{i}{2}\mathcal{P}_{+}\left\{  \left[  \varepsilon_{0}\left(
\mathbf{X}\right)  ,\mathcal{A}_{0}^{R_{l}}\right]  \mathcal{A}_{0}^{P_{l}%
}-\left[  \varepsilon_{0}\left(  \mathbf{X}\right)  ,\mathcal{A}_{0}^{P_{l}%
}\right]  \mathcal{A}_{0}^{R_{l}}\right\}  +O(\hbar^{2})
\]
which being on all point similar to the one given in \cite{Photon} or
\cite{Pierregravit1} is simply stated :
\[
M_{+}=\frac{1}{E}\left(  \left(  \frac{1}{2}\mathbf{\Sigma}-(\mathbf{A}%
_{\mathbf{R}}\mathbf{\times\tilde{p})}\right)  \mathbf{.B}-\frac{1}%
{2E}\mathbf{\nabla}\tilde{m}(\mathbf{r}).\left(  \mathbf{\tilde{p}}%
\times\mathbf{\Sigma}\right)  \right)
\]
where the \textquotedblright magnetotorsion field\textquotedblright%
\ $\mathbf{B}$ is defined through a three dimensional effective torsion tensor
$B_{\gamma}=-\frac{1}{2}P_{\delta}T^{\delta\alpha\beta}\varepsilon
_{\alpha\beta\gamma}$ where the effective torsion $T^{\delta\alpha\beta}$ is
defined as
\[
T^{\alpha\beta\delta}=\tilde{H}_{k}^{\delta}\left(  \tilde{H}^{l\alpha
}\partial_{l}\tilde{H}^{k\beta}-\tilde{H}^{l\beta}\partial_{l}\tilde
{H}^{k\alpha}\right)  +\tilde{H}^{l\alpha}\tilde{\Gamma}_{l}^{\beta\delta
}-\tilde{H}^{l\beta}\tilde{\Gamma}_{l}^{\alpha\delta}
\]
The physical origin of the term $M$ has been discussed in \cite{Photon} and
\cite{Pierregravit1} for the static case. Note that for the static case
($g^{0i}=0$) we retrieve the true torsion $T^{\alpha\beta\delta}=H_{k}%
^{\delta}\left(  H^{l\alpha}\partial_{l}H^{k\beta}-H^{l\beta}\partial
_{l}H^{k\alpha}\right)  +H^{l\alpha}\Gamma_{l}^{\beta\delta}-H^{l\beta}%
\Gamma_{l}^{\alpha\delta}.$

Considering similarly the anti-particle subspace, it turns out that the
coordinate operators are identical with the particule ones and that
$M_{-}=-M_{+}$ therefore the full energy operator with both particles and anti
particles can be cast in the form%
\begin{equation}
\varepsilon\left(  \mathbf{p,r}\right)  =\beta\widetilde{\varepsilon}+\frac
{1}{2}g^{i}p_{i}+\frac{1}{2}p_{i}g^{i}\mathbf{+}\frac{\hbar}{4E}\left(
\mathbf{\tilde{\Gamma}}^{0}+\mathbf{\Gamma}^{e}\right)  .\left(  \tilde
{m}\mathbf{\Sigma}+\frac{\left(  \mathbf{\Sigma}.\mathbf{\tilde{H}}^{i}%
P_{i}\right)  \mathbf{\tilde{H}}^{i}P_{i}}{(E+\tilde{m})}\right)  +\beta\hbar
c\frac{\left(  \left(  f-1\right)  g_{00}\left(  \mathbf{h}^{0}\times
\mathbf{h}^{i}\right)  \right)  .\left(  \nabla_{i}\mathbf{u}\right)
}{2fE\left(  1-u^{2}\right)  }\mathbf{\tilde{p}}.\mathbf{\Sigma+}\beta\hbar
M\label{Efinal}%
\end{equation}
where $M=M_{+}$ and
\[
\widetilde{\varepsilon}=c\sqrt{\left(  p_{i}+\frac{\hbar}{4E}\mathbf{\tilde
{\Gamma}}_{i}.\left(  \tilde{m}\mathbf{\Sigma}+\frac{\left(  \mathbf{\Sigma
}.\mathbf{\tilde{p}}\right)  \mathbf{\tilde{p}}}{(E+\tilde{m})}\right)
\right)  G^{ij}\left(  p_{i}+\frac{\hbar}{4E}\mathbf{\tilde{\Gamma}}%
_{i}.\left(  \tilde{m}\mathbf{\Sigma}+\frac{\left(  \mathbf{\Sigma
}.\mathbf{\tilde{p}}\right)  \mathbf{\tilde{p}}}{(E+\tilde{m})}\right)
\right)  +\tilde{m}^{2}}
\]
with the vector $\mathbf{\tilde{\Gamma}}_{i}$ defined in terms of its
components as $\left(  \mathbf{\tilde{\Gamma}}_{i}\right)  _{\gamma}%
\equiv\varepsilon_{\alpha\beta\gamma}\tilde{\Gamma}_{i}^{\alpha\beta}\left(
\mathbf{r}\right)  $. Equation $\left(  \ref{Efinal}\right)  $ is the main
result of this paper.\ 

The coordinates operators in Eq. $\left(  \ref{Efinal}\right)  $ satisfy a non
commutative algebra as
\begin{align}
\left[  r_{i},r_{j}\right]   & =i\hbar^{2}\Theta_{ij}^{rr}\\
\left[  p_{i},p_{j}\right]   & =i\hbar^{2}\Theta_{ij}^{pp}\\
\left[  p_{i},r_{j}\right]   & =-i\hbar g_{ij}+i\hbar^{2}\Theta_{ij}^{pr}%
\end{align}
where $\Theta_{ij}^{\alpha\beta}=\partial_{\alpha_{i}}\mathbf{A}_{\beta_{j}%
}-\partial_{\beta_{j}}\mathbf{A}_{\alpha_{i}}+[\mathbf{A}_{\alpha_{i}%
,}\mathbf{A}_{\beta_{j}}]$ the so-called Berry curvature. An explicit
computation gives :
\begin{align}
\Theta_{kl}^{rr}  & =\frac{-1}{2E^{3}}\left(  \tilde{m}\mathbf{\Sigma}%
_{\gamma}+\frac{\left(  \mathbf{\Sigma}^{\delta}\tilde{P}_{\delta}\right)
\tilde{P}_{\gamma}}{(E+\tilde{m})}\right)  \varepsilon^{\alpha\beta\gamma
}\tilde{H}_{\alpha}^{i}\tilde{H}_{\beta}^{j}g_{ik}g_{jl}\nonumber\\
\Theta_{ij}^{pp}  & =\frac{-1}{2E^{3}}\left(  \tilde{m}\mathbf{\Sigma}%
_{\gamma}+\frac{\left(  \mathbf{\Sigma}^{\delta}\tilde{P}_{\delta}\right)
\tilde{P}_{\gamma}}{(E+\tilde{m})}\right)  \nabla_{r_{i}}\tilde{P}_{\alpha
}\nabla_{r_{j}}\tilde{P}_{\beta}\varepsilon^{\alpha\beta\gamma}+\frac
{1}{2E^{3}}\left[  \nabla_{r_{i}}\tilde{m}\left(  \left(  \mathbf{\Sigma
}\times\mathbf{\tilde{P}}\right)  .\nabla_{r_{j}}\mathbf{\tilde{P}}\right)
-\nabla_{r_{j}}V(\mathbf{r})\left(  \left(  \mathbf{\Sigma}\times
\mathbf{\tilde{P}}\right)  .\nabla_{r_{i}}\mathbf{\tilde{P}}\right)  \right]
\nonumber\\
\Theta_{ij}^{pr}  & =\frac{1}{2E^{3}}\left(  \tilde{m}\mathbf{\Sigma}_{\gamma
}+\frac{\left(  \mathbf{\Sigma}^{\delta}\tilde{P}_{\delta}\right)  \tilde
{P}_{\gamma}}{(E+\tilde{m})}\right)  \varepsilon^{\alpha\beta\gamma}%
\nabla_{r_{i}}\tilde{P}_{\alpha}H_{\beta}^{l}g_{jl}-\frac{1}{2E^{3}}%
\nabla_{r_{i}}\tilde{m}\left(  \mathbf{\Sigma}\times\mathbf{\tilde{P}}\right)
_{j}%
\end{align}
There are also other Berry curvature mixing coordinates and spin
\begin{align}
\Theta_{ij}^{r\Sigma}  & =\left[  r_{i},\Sigma_{j}\right]  =ic^{2}\frac
{-p_{j}\mathbf{\Sigma}_{i}+\mathbf{p.\Sigma}\delta_{ij}}{E\left(  E+\tilde
{m}\right)  }\nonumber\\
\Theta_{ij}^{p\Sigma}  & =\left[  p_{i},\Sigma_{j}\right]  =-ic^{2}%
\frac{-p_{j}\mathbf{\Sigma}_{l}+\mathbf{p.\Sigma}\delta_{lj}}{E\left(
E+\tilde{m}\right)  }\tilde{H}_{l}^{\gamma}\mathbf{\nabla}_{r_{i}}\tilde
{p}_{\gamma}%
\end{align}

Interestingly Eq. $\left(  \ref{Efinal}\right)  $ can be rewritten as
\begin{align}
\varepsilon & =c\beta\sqrt{\left(  p_{i}-\frac{\hbar}{2}\mathbf{\tilde{\Gamma
}}_{i}.\mathbf{\tilde{\Theta}}^{rr}\right)  G^{ij}\left(  p_{i}-\frac{\hbar
}{2}\mathbf{\tilde{\Gamma}}_{i}.\mathbf{\tilde{\Theta}}^{rr}\right)
+\tilde{m}^{2}(\mathbf{r})}+\frac{1}{2}g^{i}p_{i}+\frac{1}{2}p_{i}%
g^{i}\mathbf{-}\frac{\hbar}{2}\left(  \mathbf{\tilde{\Gamma}}^{0}%
+\mathbf{\Gamma}^{e}\right)  .\mathbf{\tilde{\Theta}}^{rr}\nonumber\\
& +\beta\hbar c\frac{\left(  \left(  f-1\right)  g_{00}\left(  \mathbf{h}%
^{0}\times\mathbf{h}^{i}\right)  \right)  .\left(  \nabla_{i}\mathbf{u}%
\right)  }{2fE\left(  1-\mathbf{u}^{2}\right)  }\mathbf{\tilde{p}%
}.\mathbf{\Sigma}+\beta\hbar M\label{ener}%
\end{align}
where $\tilde{\Theta}_{\gamma}^{rr}=\frac{-1}{2E}\left(  \tilde{m}%
\mathbf{\Sigma}_{\gamma}+\frac{\left(  \mathbf{\Sigma}^{\delta}\tilde
{P}_{\delta}\right)  \tilde{P}_{\gamma}}{(E+\tilde{m})}\right)  $ is the
\textquotedblright rescaled\textquotedblright\ Berry curvature. This formula
clearly shows that the spin connection couples only to the Berry curvature.

We note in Eq. $\left(  \ref{ener}\right)  $ the presence of the term
$\mathbf{-}\frac{\hbar}{2}\left(  \mathbf{\tilde{\Gamma}}^{0}+\mathbf{\Gamma
}^{e}\right)  .\mathbf{\tilde{\Theta}}^{rr}+\frac{1}{2}g^{i}p_{i}+\frac{1}%
{2}p_{i}g^{i}$ not proportional to $\beta$ and independant of the particle
charge, that is discrimnating between particles and anti particles. These
terms give different energy levels for the Dirac particles and antiparticles
as $\varepsilon_{+}\neq-\varepsilon_{-}$. The coupling $\frac{-\hbar}%
{2}\left(  \mathbf{\tilde{\Gamma}}^{0}+\mathbf{\Gamma}^{e}\right)
.\mathbf{\tilde{\Theta}}^{rr}$ proportional to the spin will survive in the
case of the non-diagonal static gravitational field but cancels for a diagonal
metrics as studied in \cite{OBUKHOV}. On the other hand the term $\frac{1}%
{2}g^{i}p_{i}+\frac{1}{2}p_{i}g^{i}$ vanishes for a static metrics since in
this case $g^{0i}=\partial^{0}g^{ij}=0$ so that $g^{i}=0$. Therefore the
symmetry between particle and antiparticle is restablished only for static
diagonal metrics.

\section{The static gravitational field}

This case is caracterized by the following time independent metric:
$g_{ij}\equiv g_{ij}\left(  \mathbf{R}\right)  $, $g_{00}\equiv g_{00}\left(
\mathbf{R}\right)  $, $g_{i0}=0$. In that case expressions simplify greatly.
Actually, the transformation matrix $U$ is $U=\sqrt{-g}h_{0}^{0}$ and
$U^{\frac{1}{2}}=\sqrt{\sqrt{-g}h_{0}^{0}}=f$. The effective quantities reduce to:%

\begin{align*}
\left(  \mathbf{\tilde{H}}^{i}\right)  _{\beta}  & =\tilde{H}_{\beta}%
^{i}=H_{\beta}^{i}=g_{00}h_{0}^{0}h_{\beta}^{i}=\sqrt{g_{00}}h_{\beta}^{i}\\
\tilde{\Gamma}_{i}^{\varrho\beta}  & =\hat{\Gamma}_{i}^{\varrho\beta}%
=\Gamma_{i}^{\varrho\beta}\\
\hat{\Gamma}_{\gamma}^{0}  & =\tilde{\Gamma}_{\gamma}^{0}=\frac{1}{4}\left\{
\varepsilon_{\varrho\beta\gamma}\Gamma_{0}^{\varrho\beta}+\varepsilon_{\text{
}\nu\gamma}^{\beta}H_{\beta}^{i}\Gamma_{i}^{0\nu}\right\}  =-\frac{1}{4}%
g_{00}\left(  h_{0}^{0}\right)  ^{2}\varepsilon_{\text{ }\nu\gamma}^{\beta
}h_{\beta}^{i}h_{l}^{\nu}h_{i}^{\lambda}\Gamma_{\lambda}^{kl}\\
\mathbf{\Gamma}^{e}  & =0\\
G^{ij}  & =g_{00}g^{ij}\\
\tilde{m}  & =mg_{00}h_{0}^{0}\text{ as }u^{\beta}=0
\end{align*}
where $\Gamma_{\lambda}^{kl}$ stands for the Christoffel symbol. In this case
Eq. $\left(  \ref{Efinal}\right)  $ reduces to%

\begin{align}
\varepsilon & \simeq c\beta\sqrt{\left(  p_{i}+\frac{\hbar}{4E}\mathbf{\Gamma
}_{i}\left(  \mathbf{r}\right)  .\left(  \widetilde{m}\mathbf{\Sigma}%
+\frac{\left(  \mathbf{\Sigma}.\mathbf{\tilde{p}}\right)  \mathbf{\tilde{p}}%
}{(E+\widetilde{m})}\right)  \right)  G^{ij}\left(  \mathbf{r}\right)  \left(
p_{i}+\frac{\hbar}{4E}\mathbf{\Gamma}_{i}\left(  \mathbf{r}\right)  .\left(
\widetilde{m}\mathbf{\Sigma}+\frac{\left(  \mathbf{\Sigma}.\mathbf{\tilde{p}%
}\right)  \mathbf{\tilde{p}}}{(E+\widetilde{m})}\right)  \right)
+\widetilde{m}^{2}}+\hbar\beta M\nonumber\\
& \mathbf{+}\frac{\hbar}{4E}\tilde{\Gamma}_{\gamma}^{0}\left(  \widetilde
{m}\mathbf{\Sigma}^{\gamma}+\frac{\left(  \mathbf{\Sigma}.\mathbf{\tilde{p}%
}\right)  \tilde{p}^{\gamma}}{(E+\widetilde{m})}\right)
\end{align}
with $\tilde{p}_{\alpha}\mathbf{=}\sqrt{g_{00}}h_{\alpha}^{i}(p_{i}%
+\frac{\hbar}{4}\varepsilon_{\varrho\beta\gamma}\Gamma_{i}^{\varrho\beta
}\sigma^{\gamma})$ and the vector $\mathbf{\Gamma}_{i}$ defined in terms of
its components as $\left(  \mathbf{\Gamma}_{i}\right)  _{\gamma}%
\equiv\varepsilon_{\alpha\beta\gamma}\Gamma_{i}^{\alpha\beta}\left(
\mathbf{r}\right)  .$ Then even for this case, particles and antiparticles
avec a different in-band energy operator because of term $\frac{\hbar}%
{4E}\tilde{\Gamma}_{\gamma}^{0}\left(  \widetilde{m}\mathbf{\Sigma}^{\gamma
}+\frac{\left(  \mathbf{\Sigma}.\mathbf{\tilde{p}}\right)  \tilde{p}^{\gamma}%
}{(E+\widetilde{m})}\right)  $ breaking the symmetry $\varepsilon_{+}%
\neq-\varepsilon_{-}.$ Although this term was already in previous studies
\cite{Photon}\cite{Pierregravit1} this essential point has not been pointed
out.\ For a diagonal metric $\mathbf{\tilde{\Gamma}}_{0}=0$, and the symmetry
particles/anti-particles is recovered.

\section{Ultrarelativistic limit}

It is interesting to look at the ultrarelativistic limit $mc^{2}\rightarrow0
$. One readily obtain
\begin{equation}
\varepsilon\simeq\beta\widetilde{\varepsilon}+c\beta\frac{\left(  \left(
f-1\right)  g_{00}\left(  \mathbf{h}^{0}\times\mathbf{h}^{i}\right)  \right)
.\left(  \nabla_{i}\mathbf{u}\right)  }{2f\left(  1-\mathbf{u}^{2}\right)
}\tilde{\lambda}+\beta\frac{\tilde{\lambda}g_{00}}{2E}\frac{\mathbf{B}%
.\mathbf{\tilde{p}}}{\tilde{p}}+\frac{\tilde{\lambda}}{4}\frac{\tilde
{p}^{\gamma}\left(  \tilde{\Gamma}_{\gamma}^{0}+\mathbf{\Gamma}_{\gamma}%
^{e}\right)  }{\tilde{p}}+\frac{1}{2}g^{i}p_{i}+\frac{1}{2}p_{i}%
g^{i}\label{phot}%
\end{equation}
with $\widetilde{\varepsilon}=c\sqrt{\left(  p_{i}+\frac{\tilde{\lambda}}%
{4}\frac{\tilde{\Gamma}_{i}(\mathbf{r}).\mathbf{p}}{p}\right)  G^{ij}\left(
p_{j}+\frac{\tilde{\lambda}}{4}\frac{\tilde{\Gamma}_{j}(\mathbf{r}%
).\mathbf{p}}{p}\right)  }$ and $\tilde{\lambda}=\hbar\mathbf{\tilde{p}%
.\Sigma/}\tilde{p}$ a biased helicity, that is not projected on the momentum
$\mathbf{p}$ but rather on $\mathbf{\tilde{p}}$. This fact is not astonishing.
Actually, as we saw in the diagonalization process, the particle is submitted
to the action of an effective gravitationnal field, which differs slightly
from the initial field. This effective metric is responsible for considering
the momentum $\mathbf{\tilde{p}}$ rather than as a dynamical variable
$\mathbf{P} $. Nicely this energy can be expressed in terms of the helicity
which is the relevant variable for massless particles and not in term of
$\Sigma$. As shown in \cite{Photon}, Eq. $\left(  \ref{phot}\right)  $ is also
valid for photon with the one-half spin matrix $\mathbf{\Sigma}$ replaced with
spin one matrix $\mathbf{S.}$

Here also we see that photons and anti-photons do not have the same energy
spectrum. The symmetry is again only restored for a static diagonal metric.

\section{The time dependent symmetric gravitational field}

A typical example of such a metric is the Schwarzschild space-time in
isotropic coordinates. This case, studied in a different manner in
\cite{OBUKHOV}and \cite{SILENKO} for a time independent metric, received a
full independent treatment within our formalism in \cite{Photon}%
\cite{Pierregravit1}. We now present the equivalent results for a time
dependent metric, completed with the spin matrix dynamics. For a symmetric
metric, the semiclassical Hamiltonian has the following form \cite{SILENKO}%

\begin{equation}
H_{0}=\frac{1}{2}\left(  \alpha.PF(\mathbf{R},t)+F(\mathbf{R},t)\alpha
.P\right)  +\beta mV(\mathbf{R},t)
\end{equation}
corresponding to the metric $g_{ij}=\delta_{ij}\left(  \frac{V(\mathbf{R}%
,t)}{F(\mathbf{R})}\right)  ^{2}$, $g_{i0}=0$ and $g_{00}=V^{2}(\mathbf{R}%
,t)$. We will define $\phi=\frac{V}{F}$. In that context the relevant
quantities for the diagonalization appear to be :%
\begin{align*}
h_{i}^{\beta}  & =\phi\delta_{i}^{\beta}\text{, \ \ }h_{\beta}^{i}=\frac
{1}{\phi}\delta_{\beta}^{i}\\
h_{0}^{\beta}  & =V(\mathbf{R},t)\delta_{0}^{\beta}\text{, \ \ }h_{\beta}%
^{0}=\frac{1}{V(\mathbf{R},t)}\delta_{\beta}^{0}%
\end{align*}
and

\
\begin{align*}
\left(  \mathbf{H}^{i}\right)  _{\beta}  & =H_{\beta}^{i}=g_{00}h_{0}%
^{0}h_{\beta}^{i}=Vh_{\beta}^{i}\\
\left(  \mathbf{\tilde{H}}^{i}\right)  _{\beta}  & =\tilde{H}_{\beta}%
^{i}=H_{\beta}^{i}\\
\tilde{\Gamma}_{i}^{\varrho\beta}  & =\hat{\Gamma}_{i}^{\varrho\beta}%
=\Gamma_{i}^{\varrho\beta}=\frac{\left(  \partial^{\rho}\phi h_{i}^{\beta
}-\partial^{\beta}\phi h_{i}^{\rho}\right)  }{\phi^{2}}\\
\Gamma_{i}^{0\nu}  & =\hbar\frac{\partial^{0}\phi h_{i}^{\nu}}{\phi V}\\
\tilde{\Gamma}_{\gamma}^{0}  & =\hat{\Gamma}_{\gamma}^{0}=\frac{\hbar}%
{4}\varepsilon_{\varrho\beta\gamma}\frac{F(\mathbf{R})}{V^{2}(\mathbf{R}%
)}\partial^{0}g^{\varrho\beta}+\hbar g_{00}h_{0}^{0}\frac{F(\mathbf{R})}%
{V^{2}(\mathbf{R})}\left(  \partial^{0}\left(  \frac{V(\mathbf{R}%
)}{F(\mathbf{R})}\right)  \right)  \varepsilon_{\text{ }\nu\gamma}^{\beta
}h_{\beta}^{i}h_{i}^{\nu}=0\\
\mathbf{\Gamma}^{e}  & =0\\
G^{ij}  & =V^{2}g^{ij}%
\end{align*}
Similar computations to the ones performed in the previous section lead to the
following expressions for the dynamical variables and the diagonalized
Hamiltonian
\begin{equation}
\mathbf{r}=\mathbf{R-}\hbar\frac{F^{2}(\mathbf{R},t)\mathbf{\Sigma}%
\times\mathbf{P}}{2E(E+mV(\mathbf{R}))},\text{ \ \ }\mathbf{p}=\mathbf{P}%
\end{equation}
and the diagonal energy becomes:%
\begin{equation}
\varepsilon=\beta\sqrt{F^{2}(\mathbf{r,}t)\mathbf{P}^{2}+\mathbf{P}^{2}%
F^{2}(\mathbf{r,}t)+mV^{2}(\mathbf{r,}t)}-\frac{F^{3}(\mathbf{r,}t)}{2E^{2}%
}m\hbar\beta\mathbf{\nabla}\phi(\mathbf{r,}t).\left(  \mathbf{P}%
\times\mathbf{\Sigma}\right)
\end{equation}
The Berry curvatures are given by:
\begin{align}
\Theta_{ij}^{rr}  & =-\frac{\hbar F^{3}(\mathbf{r},t)\varepsilon^{ijk}}%
{2E^{3}}\left(  m\phi(\mathbf{r},t)\mathbf{\Sigma}_{k}+\frac{F(\mathbf{r}%
,t)\left(  \mathbf{\Sigma.P}\right)  \mathbf{P}_{k}}{E+mV(\mathbf{r})}\right)
\\
\Theta_{ij}^{pr}  & =-\frac{\hbar F^{3}(\mathbf{r},t)}{2E^{3}}m\nabla_{i}%
\phi(\mathbf{r})\left(  \mathbf{\Sigma}\times\mathbf{P}\right)  _{j}\\
\Theta_{ij}^{pp}  & =0
\end{align}
and $\Theta_{ij}^{r\Sigma}$ being unchanged, $\Theta_{ij}^{p\Sigma}=0$, and
$\widetilde{\varepsilon}=\sqrt{F^{2}(\mathbf{r,}t)\mathbf{P}^{2}%
+\mathbf{P}^{2}F^{2}(\mathbf{r,}t)+mV^{2}(\mathbf{r,}t)}.$

Note here that the magnetotorsion field $B=0$. From Appendix B, we can also
get the non Hermitian contributions the time evolution operator which in this
case reads $i\hbar\frac{\partial}{\partial t}\ln f=\frac{1}{2}\frac{\partial
}{\partial t}\ln\sqrt{-g}h_{0}^{0}$.

One can check, after developing $\mathbf{r}$ as a function of $\mathbf{R}$ and
the Berry phase, that our Hamiltonian coincides, in the weak field
approximation,with the one given in \cite{SILENKO} when considering the
semiclassical limit (order $\hbar$). This also confirms the validity of the
Foldy Wouthuysen approach asserted in \cite{SILENKO}.

\section{Only time dependent metric}

The metric tensor and the vierbein only depend on time. Therefore we have%

\begin{align*}
\left(  \mathbf{H}^{i}\right)  _{\beta}  & =H_{\beta}^{i}=g_{00}h_{0}%
^{0}\left(  h_{\beta}^{i}-\frac{h_{\beta}^{0}h_{0}^{i}}{h_{0}^{0}}\right) \\
\left(  \mathbf{\tilde{H}}^{i}\right)  _{\beta}  & =\tilde{H}_{\beta}%
^{i}=H_{\beta}^{i}+\frac{2g_{00}\left(  h_{\delta}^{0}h_{\beta}^{i}-h_{\beta
}^{0}h_{\delta}^{i}\right)  u_{\delta}}{\left(  1-\mathbf{u}^{2}\right)  }\\
\hat{\Gamma}_{i}^{\varrho\beta}  & =\Gamma_{i}^{\varrho\beta}-\hbar
\varepsilon_{\text{ \ \ }\gamma}^{\varrho\beta}\left(  H^{-1}\right)
_{i}^{\kappa}g_{00}h_{\delta}^{0}h_{\eta}^{i}\varepsilon_{\text{ \ \ \ }%
\kappa}^{\delta\eta}\frac{\Gamma_{j}^{0\gamma}}{4}\\
\tilde{\Gamma}_{i}^{\varrho\beta}  & =\left(  \tilde{H}^{-1}\right)
_{i}^{\eta}H_{\eta}^{j}\hat{\Gamma}_{j}^{\varrho\beta}\\
\hat{\Gamma}_{\gamma}^{0}  & =\frac{1}{4}\left(  \varepsilon_{\varrho
\beta\gamma}\Gamma_{0}^{\varrho\beta}+g_{00}g^{0i}\varepsilon_{\varrho
\beta\gamma}\Gamma_{i}^{\varrho\beta}+\varepsilon_{\text{ }\nu\gamma}^{\beta
}H_{\beta}^{i}\Gamma_{i}^{0\nu}\right) \\
\tilde{\Gamma}_{\gamma}^{0}  & =\frac{\left(  1+\mathbf{u}^{2}\right)
\delta_{\gamma}^{\eta}-u^{\eta}u_{\gamma}}{\left(  1-\mathbf{u}^{2}\right)
}\hat{\Gamma}_{\eta}^{0}+\left(  \left(  H_{\delta}^{i}\mathbf{\hat{\Gamma}%
}_{i}^{\delta}\mathbf{+}\frac{\mathbf{\Gamma}_{0}^{0}}{2}\mathbf{-}%
g_{00}g^{0i}\frac{\mathbf{\Gamma}_{i}^{0}}{2}\right)  \times\mathbf{u}\right)
_{\gamma}\\
\mathbf{\Gamma}_{\gamma}^{e}  & =0
\end{align*}
so that the energy becomes
\begin{equation}
\varepsilon\left(  \mathbf{p,r,}t\right)  =\beta\tilde{\varepsilon}%
\mathbf{+}\frac{\hbar}{4E}\mathbf{\tilde{\Gamma}}^{0}.\left(  \tilde
{m}\mathbf{\Sigma}+\frac{\left(  \mathbf{\Sigma}.\mathbf{\tilde{H}}^{i}%
p_{i}\right)  \mathbf{\tilde{H}}^{i}p_{i}}{(E+\tilde{m})}\right)  +\frac{1}%
{2}g^{i}(t)p_{i}+\frac{1}{2}p_{i}g^{i}(t)\mathbf{+}\beta\hbar M
\end{equation}
with
\[
\tilde{\varepsilon}=c\sqrt{\left(  p_{i}+\frac{\hbar}{4E}\mathbf{\Gamma}%
_{i}\left(  \mathbf{r},t\right)  .\left(  \widetilde{m}\mathbf{\Sigma}%
+\frac{\left(  \mathbf{\Sigma}.\mathbf{\tilde{p}}\right)  \mathbf{\tilde{p}}%
}{(E+\widetilde{m})}\right)  \right)  G^{ij}\left(  \mathbf{r},t\right)
\left(  p_{i}+\frac{\hbar}{4E}\mathbf{\Gamma}_{i}\left(  \mathbf{r},t\right)
.\left(  \widetilde{m}\mathbf{\Sigma}+\frac{\left(  \mathbf{\Sigma
}.\mathbf{\tilde{p}}\right)  \mathbf{\tilde{p}}}{(E+\widetilde{m})}\right)
\right)  +\widetilde{m}^{2}}
\]
which show that particles and antiparticles have a different energy spectrum
in this graviational field.

\section{Conclusion}

The semiclassical limit for Dirac particles interacting with a fully general
gravitational field was investigated through a first order in $\hbar$
diagonalization of the Dirac Hamiltonian. This work extends previous ones
where only static metrics were considered. The time dependence of the metrics
leads to new contributions of the in-band energy operator. In particular we
found a coupling term between the linear momentum and the spin, and terms
which in general will break the particle - antiparticle symmetry.

As already found by other authors, the time dependence leads also to special
features like the non-unitarity of the evolution operator, whose origin can be
tracked back to the notion of scalar product in the Hilbert space of wave
functions for a time dependent metric. This non-unitarity is unavoidable but
we could nevertheless diagonalize the full evolution operator, even though our
main focus was to obtain the block-diagonal form of the energy, this one
turning out to be Hermitian. In addition, to the very general semiclassical
diagonal energy operator, we provided several physically relevant examples.

\section{Appendix.}

\subsection{Derivation of Eq. $\left(  \ref{HH}\right)  $.}

Let us start with the following development for $\hat{H}$:%

\begin{align*}
\hat{H}  & =g_{00}h_{\beta}^{0}\gamma^{\beta}\gamma^{\alpha}\hat{P}_{\alpha
}+\frac{\hbar}{4}\varepsilon_{\varrho\beta\gamma}\Gamma_{0}^{\varrho\beta
}\mathbf{\Sigma}^{\gamma}+i\frac{\hbar}{4}\Gamma_{0}^{0\beta}\alpha^{\beta
}\text{\ }+g_{00}h_{\beta}^{0}\gamma^{\beta}m+\frac{i}{2}\partial_{t}%
\ln\left(  -gg^{00}\right) \\
& =g_{00}g^{0i}(P_{i}+\hbar\varepsilon_{\varrho\beta\gamma}\frac{\Gamma
_{i}^{\varrho\beta}}{4}\mathbf{\Sigma}^{\gamma}+i\hbar\frac{\Gamma
_{i}^{0\gamma}}{4}\alpha^{\gamma})\\
& +ig_{00}\left(  h_{\delta}^{0}h_{\eta}^{i}\varepsilon^{\delta\eta\kappa
}\right)  \mathbf{\Sigma}_{\kappa}(P_{i}+\hbar\varepsilon_{\varrho\beta\gamma
}\frac{\Gamma_{i}^{\varrho\beta}}{4}\mathbf{\Sigma}^{\gamma}+i\hbar
\frac{\Gamma_{i}^{0\gamma}}{4}\alpha^{\gamma})\\
& +g_{00}h_{0}^{0}\alpha^{\beta}h_{\beta}^{i}(P_{i}+\hbar\varepsilon
_{\varrho\beta\gamma}\frac{\Gamma_{i}^{\varrho\beta}}{4}\mathbf{\Sigma
}^{\gamma}+i\hbar\frac{\Gamma_{i}^{0\gamma}}{4}\alpha^{\gamma})\\
& -g_{00}\alpha^{\beta}h_{\beta}^{0}h_{0}^{i}(P_{i}+\hbar\varepsilon
_{\varrho\beta\gamma}\frac{\Gamma_{i}^{\varrho\beta}}{4}\mathbf{\Sigma
}^{\gamma}+i\hbar\frac{\Gamma_{i}^{0\gamma}}{4}\alpha^{\gamma})\\
& +\frac{\hbar}{4}\varepsilon_{\varrho\beta\gamma}\Gamma_{0}^{\varrho\beta
}\mathbf{\Sigma}^{\gamma}+i\frac{\hbar}{4}\Gamma_{0}^{0\beta}\alpha^{\beta
}\text{\ }+g_{00}h_{\beta}^{0}\gamma^{\beta}m+\frac{i}{2}\partial_{t}%
\ln\left(  -gg^{00}\right)
\end{align*}
and remark that we can rewrite the four first terms in the following form :
\begin{align*}
& g_{00}h_{0}^{0}\alpha^{\beta}\left(  h_{\beta}^{i}-\frac{h_{\beta}^{0}%
h_{0}^{i}}{h_{0}^{0}}\right)  (P_{i}+\hbar\varepsilon_{\varrho\beta\gamma
}\frac{\Gamma_{i}^{\varrho\beta}}{4}\mathbf{\Sigma}^{\gamma}+i\hbar
\frac{\Gamma_{i}^{0\gamma}}{4}\alpha^{\gamma})\\
& +g_{00}g^{0i}(P_{i}+\hbar\varepsilon_{\varrho\beta\gamma}\frac{\Gamma
_{i}^{\varrho\beta}}{4}\mathbf{\Sigma}^{\gamma}+i\hbar\frac{\Gamma
_{i}^{0\gamma}}{4}\alpha^{\gamma})\\
& +ig_{00}\left(  h_{\delta}^{0}h_{\eta}^{i}\varepsilon^{\delta\eta\kappa
}\right)  \mathbf{\Sigma}_{\kappa}(P_{i}+\hbar\varepsilon_{\varrho\beta\gamma
}\frac{\Gamma_{i}^{\varrho\beta}}{4}\mathbf{\Sigma}^{\gamma})\\
& -\hbar g_{00}\left(  h_{\delta}^{0}h_{\eta}^{i}\varepsilon^{\delta\eta
\kappa}\right)  \alpha_{\kappa}\frac{\Gamma_{i}^{0\gamma}}{4}\mathbf{\Sigma
}^{\gamma}\\
& =g_{00}h_{0}^{0}\alpha^{\beta}\left(  h_{\beta}^{i}-\frac{h_{\beta}^{0}%
h_{0}^{i}}{h_{0}^{0}}\right) \\
& \times\left(  P_{i}+\hbar\varepsilon_{\varrho\beta\gamma}\frac{\Gamma
_{i}^{\varrho\beta}}{4}\mathbf{\Sigma}^{\gamma}-\left(  \left(  h_{\beta}%
^{i}-\frac{h_{\beta}^{0}h_{0}^{i}}{h_{0}^{0}}\right)  ^{-1}\right)
_{i}^{\kappa}\left(  \frac{\hbar}{h_{0}^{0}}\left(  h_{\delta}^{0}h_{\eta}%
^{j}\varepsilon_{\text{ \ \ }\kappa}^{\delta\eta}\right)  \right)
\frac{\Gamma_{j}^{0\gamma}}{4}\mathbf{\Sigma}^{\gamma}+i\hbar\frac{\Gamma
_{i}^{0\gamma}}{4}\alpha^{\gamma}\right) \\
& +g_{00}g^{0i}(P_{i}+\hbar\varepsilon_{\varrho\beta\gamma}\frac{\Gamma
_{i}^{\varrho\beta}}{4}\mathbf{\Sigma}^{\gamma}+i\hbar\frac{\Gamma
_{i}^{0\gamma}}{4}\alpha^{\gamma})\\
& +ig_{00}\left(  h_{\delta}^{0}h_{\eta}^{i}\varepsilon^{\delta\eta\kappa
}\right)  \mathbf{\Sigma}_{\kappa}(P_{i}+\hbar\varepsilon_{\varrho\beta\gamma
}\frac{\Gamma_{i}^{\varrho\beta}}{4}\mathbf{\Sigma}^{\gamma})\\
& =\alpha^{\beta}H_{\beta}^{i}\left(  P_{i}+\hbar\varepsilon_{\varrho
\beta\gamma}\frac{\tilde{\Gamma}_{i}^{\varrho\beta}}{4}\mathbf{\Sigma}%
^{\gamma}+i\hbar\frac{\Gamma_{i}^{0\gamma}}{4}\alpha^{\gamma}\right)
+g_{00}g^{0i}(P_{i}+\hbar\varepsilon_{\varrho\beta\gamma}\frac{\Gamma
_{i}^{\varrho\beta}}{4}\mathbf{\Sigma}^{\gamma}+i\hbar\frac{\Gamma
_{i}^{0\gamma}}{4}\alpha^{\gamma})\\
& +ig_{00}\left(  h_{\delta}^{0}h_{\eta}^{i}\varepsilon^{\delta\eta\kappa
}\right)  \mathbf{\Sigma}_{\kappa}(P_{i}+\hbar\varepsilon_{\varrho\beta\gamma
}\frac{\Gamma_{i}^{\varrho\beta}}{4}\mathbf{\Sigma}^{\gamma})
\end{align*}
where the effective dreibein and spin connection :
\begin{align*}
H_{\beta}^{i}  & =g_{00}h_{0}^{0}\left(  h_{\beta}^{i}-\frac{h_{\beta}%
^{0}h_{0}^{i}}{h_{0}^{0}}\right) \\
\hat{\Gamma}_{i}^{\varrho\beta}  & =\Gamma_{i}^{\varrho\beta}-\hbar
\varepsilon_{\text{ \ \ }\gamma}^{\varrho\beta}\left(  H^{-1}\right)
_{i}^{\kappa}g_{00}\left(  h_{\delta}^{0}h_{\eta}^{j}\varepsilon_{\text{
\ \ \ }\kappa}^{\delta\eta}\right)  \frac{\Gamma_{j}^{0\gamma}}{4}%
\end{align*}
have the form claimed in the text.

Now, we compute $\hat{H}+\hat{H}^{+}$, which is the first contribution in
Eq.$\left(  \ref{dev}\right)  $ (up to the factors $\frac{1}{\left(
1-u_{\beta}u^{\beta}\right)  }$ that will be skipped constantly in this
section for the sake of readability, and reintroduced ultimately) .%

\begin{align*}
& \frac{1}{2}\left(  \hat{H}+\hat{H}^{+}\right)  =\frac{1}{2}\alpha^{\beta
}H_{\beta}^{i}\left(  P_{i}+\hbar\varepsilon_{\varrho\beta\gamma}\frac
{\hat{\Gamma}_{i}^{\varrho\beta}}{4}\mathbf{\Sigma}^{\gamma}\right)  +\frac
{1}{2}\left(  P_{i}+\hbar\varepsilon_{\varrho\beta\gamma}\frac{\hat{\Gamma
}_{i}^{\varrho\beta}}{4}\mathbf{\Sigma}^{\gamma}\right)  \alpha^{\beta
}H_{\beta}^{i}\\
& +\frac{1}{2}g_{00}g^{0i}P_{i}+\frac{1}{2}P_{i}g_{00}g^{0i}\\
& +g_{00}g^{0i}(\hbar\varepsilon_{\varrho\beta\gamma}\frac{\Gamma_{i}%
^{\varrho\beta}}{4}\mathbf{\Sigma}^{\gamma})-\hbar\nabla_{R_{i}}\left(
g_{00}\left(  h_{\delta}^{0}h_{\eta}^{i}\varepsilon^{\delta\eta\kappa}\right)
\right)  \mathbf{\Sigma}_{\kappa}-g_{00}\left(  h_{\delta}^{0}h_{\eta}%
^{i}\varepsilon^{\delta\eta\kappa}\right)  \varepsilon_{\kappa}^{\text{
}\gamma\nu}(\hbar\varepsilon_{\varrho\beta\gamma}\frac{\Gamma_{i}%
^{\varrho\beta}}{4})\mathbf{\Sigma}_{\nu}\\
& -\hbar\varepsilon_{\text{ }\gamma\nu}^{\beta}H_{\beta}^{i}\frac{\Gamma
_{i}^{0\gamma}}{4}\mathbf{\Sigma}^{\nu}+\frac{\hbar}{4}\varepsilon
_{\varrho\beta\gamma}\Gamma_{0}^{\varrho\beta}\mathbf{\Sigma}^{\gamma}%
+g_{00}h_{0}^{0}\beta m\\
& =\frac{1}{2}\alpha^{\beta}H_{\beta}^{i}\left(  P_{i}+\hbar\varepsilon
_{\varrho\beta\gamma}\frac{\hat{\Gamma}_{i}^{\varrho\beta}}{4}\mathbf{\Sigma
}^{\gamma}\right)  +\frac{1}{2}\left(  P_{i}+\hbar\varepsilon_{\varrho
\beta\gamma}\frac{\hat{\Gamma}_{i}^{\varrho\beta}}{4}\mathbf{\Sigma}^{\gamma
}\right)  \alpha^{\beta}H_{\beta}^{i}+g_{00}h_{0}^{0}\beta m\\
& +\left(  \mathbf{\hat{\Gamma}}^{0}-\hbar\nabla_{R_{i}}\left(  g_{00}\left(
h_{\delta}^{0}h_{\eta}^{i}\varepsilon_{\text{ \ \ }\gamma}^{\delta\eta
}\right)  \right)  \right)  .\mathbf{\Sigma}+\frac{1}{2}g_{00}g^{0i}%
P_{i}+\frac{1}{2}P_{i}g_{00}g^{0i}%
\end{align*}
with :
\begin{align*}
\left(  \mathbf{\hat{\Gamma}}^{0}\right)  _{\gamma}  & =\frac{\hbar}%
{4}\varepsilon_{\varrho\beta\gamma}\Gamma_{0}^{\varrho\beta}+\hbar
g_{00}g^{0i}\varepsilon_{\varrho\beta\gamma}\frac{\Gamma_{i}^{\varrho\beta}%
}{4}\\
& +\hbar g_{00}\left(  h_{\delta}^{0}h_{\eta}^{i}\varepsilon^{\delta\eta
\kappa}\right)  \varepsilon_{\kappa\gamma}^{\text{ \ \ }\nu}(\varepsilon
_{\varrho\beta\nu}\frac{\Gamma_{i}^{\varrho\beta}}{4})+\hbar\varepsilon
_{\text{ }\nu\gamma}^{\beta}H_{\beta}^{i}\frac{\Gamma_{i}^{0\nu}}{4}%
\end{align*}
Introduce, as in the text $\left(  \mathbf{\Gamma}^{0}\right)  _{\gamma
}=\Gamma_{0\gamma}=\varepsilon_{\varrho\beta\gamma}\Gamma_{0}^{\varrho\beta}$,
$\left(  \mathbf{\Gamma}^{i}\right)  _{\gamma}=g_{00}g^{0i}\hbar
\varepsilon_{\varrho\beta\gamma}\Gamma_{i}^{\varrho\beta}$, and use the
following expression for the spin connection
\[
\Gamma_{i}^{\varrho\beta}=h_{\mu}^{\varrho}h_{\nu}^{\beta}\left(  h_{i}^{\xi
}\Gamma_{i}^{\mu\nu}-\nabla^{\mu}h_{i}^{\nu}\right)
\]
to show that $\hbar g_{00}\left(  h_{\delta}^{0}h_{\eta}^{i}\varepsilon
^{\delta\eta\kappa}\right)  \varepsilon_{\kappa\gamma}^{\text{ \ \ }\nu
}(\varepsilon_{\varrho\beta\nu}\frac{\Gamma_{i}^{\varrho\beta}}{4})=0$. As a
consequence, we are left with%
\[
\mathbf{\hat{\Gamma}}^{0}=\frac{\hbar}{4}\left(  \mathbf{\Gamma}_{0}%
+g_{00}g^{0i}\mathbf{\Gamma}_{i}\right)  +\hbar\mathbf{H}^{i}\times
\frac{\mathbf{\Gamma}_{i}^{0}}{4}
\]
as announced.

To compute the other contributions to $U^{\frac{1}{2}}\hat{H}U^{-\frac{1}{2}}
$ in Eq. $\left(  \ref{dev}\right)  $, we need to find the expressions for
\ $-u_{\beta}\alpha^{\beta}\frac{\left(  H+H^{+}\right)  }{2}u_{\beta}%
\alpha^{\beta}$, $\frac{-\left[  H,u_{\beta}\alpha^{\beta}\right]  +\left[
H^{+},u_{\beta}\alpha^{\beta}\right]  }{2}$, and $-\frac{i\left(  1+u_{\beta
}\alpha^{\beta}\right)  \mathbf{\nabla}_{\mathbf{P}}H\mathbf{\nabla
}_{\mathbf{R}}\left(  u_{\beta}u^{\beta}\right)  }{2\left(  \left(
1-u_{\beta}u^{\beta}\right)  \right)  ^{2}}\left(  1-u_{\beta}\alpha^{\beta
}\right)  $. We start with the computation of $u_{\beta}\alpha^{\beta}\left(
\hat{H}+\hat{H}^{+}\right)  u_{\beta}\alpha^{\beta}$ by decomposing $\left(
\hat{H}+\hat{H}^{+}\right)  $ as:
\begin{align*}
\frac{\left(  \hat{H}+\hat{H}^{+}\right)  }{2}  & =\frac{\left(  \hat
{H}^{\prime}+\hat{H}^{\prime+}\right)  }{2}+\left(  \hat{\Gamma}_{\gamma}%
^{0}-\hbar\nabla_{R_{i}}\left(  g_{00}\left(  \mathbf{h}^{0}\times
\mathbf{h}^{i}\right)  \right)  \right)  \Sigma^{\gamma}+g_{00}h_{0}^{0}\beta
m\\
& +\frac{1}{2}g_{00}g^{0i}P_{i}+\frac{1}{2}P_{i}g_{00}g^{0i}%
\end{align*}
where
\[
\frac{\left(  \hat{H}^{\prime}+\hat{H}^{\prime+}\right)  }{2}=\frac{1}%
{2}\alpha^{\beta}H_{\beta}^{i}\left(  P_{i}+\hbar\varepsilon_{\varrho
\beta\gamma}\frac{\hat{\Gamma}_{i}^{\varrho\beta}}{4}\mathbf{\Sigma}^{\gamma
}\right)  +\frac{1}{2}\left(  P_{i}+\hbar\varepsilon_{\varrho\beta\gamma}%
\frac{\hat{\Gamma}_{i}^{\varrho\beta}}{4}\mathbf{\Sigma}^{\gamma}\right)
\alpha^{\beta}H_{\beta}^{i}
\]
For the sake of the computations, we will denote $\hat{P}_{\alpha}=H_{\alpha
}^{i}\left(  P_{i}+\hbar\varepsilon_{\varrho\beta\gamma}\frac{\hat{\Gamma}%
_{i}^{\varrho\beta}}{4}\mathbf{\Sigma}^{\gamma}\right)  \equiv H_{\alpha}%
^{i}\hat{P}_{i}$.so that%

\begin{align*}
u_{\beta}\alpha^{\beta}\frac{\left(  \hat{H}^{\prime}+\hat{H}^{\prime
+}\right)  }{2}u_{\beta}\alpha^{\beta}  & =\frac{1}{2}u_{\beta}\alpha^{\beta
}\alpha^{\beta}H_{\beta}^{i}\hat{P}^{i}u_{\beta}\alpha^{\beta}+\frac{1}%
{2}u_{\beta}\alpha^{\beta}\hat{P}^{i}\alpha^{\beta}u_{\beta}\alpha^{\beta
}H_{\beta}^{i}\\
& =\frac{1}{2}u_{\gamma}\alpha^{\gamma}\alpha^{\beta}H_{\beta}^{i}\hat{P}%
^{i}u_{\gamma}\alpha^{\gamma}+\frac{1}{2}u_{\gamma}\alpha^{\gamma}\hat{P}%
^{i}\alpha^{\beta}u_{\gamma}\alpha^{\gamma}H_{\beta}^{i}\\
& =\frac{1}{2}\sqrt{g_{00}}u_{\gamma}\alpha^{\gamma}\alpha^{\beta}u_{\gamma
}\alpha^{\gamma}H_{\beta}^{i}\hat{P}^{i}+\frac{1}{2}\hat{P}^{i}H_{\beta}%
^{i}u_{\gamma}\alpha^{\gamma}\alpha^{\beta}u_{\gamma}\alpha^{\gamma}\\
& +\frac{1}{2}u_{\gamma}\alpha^{\gamma}\alpha^{\beta}H_{\beta}^{i}\left[
\hat{P}^{i},u_{\gamma}\alpha^{\gamma}\right]  +\frac{1}{2}\left[  u_{\gamma
}\alpha^{\gamma},\hat{P}^{i}\right]  \alpha^{\beta}u_{\gamma}\alpha^{\gamma
}H_{\beta}^{i}\sqrt{g_{00}}\\
& =\frac{1}{2}u_{\gamma}\alpha^{\gamma}\alpha^{\beta}u_{\gamma}\alpha^{\gamma
}H_{\beta}^{i}\hat{P}^{i}+\frac{1}{2}\hat{P}^{i}H_{\beta}^{i}u_{\gamma}%
\alpha^{\gamma}\alpha^{\beta}u_{\gamma}\alpha^{\gamma}\\
& +\frac{1}{2}iu_{\gamma}\varepsilon^{\gamma\beta\delta}\left(  \left[
\mathbf{\Sigma}^{\delta}H_{\beta}^{i}\left[  \hat{P}^{i},u_{\gamma}%
\alpha^{\gamma}\right]  +H_{\beta}^{i}\left[  \hat{P}^{i},u_{\gamma}%
\alpha^{\gamma}\right]  \mathbf{\Sigma}^{\delta}\right]  \right) \\
& =\frac{1}{2}\alpha^{\beta}\mathbf{u}^{2}H_{\beta}^{i}\hat{P}^{i}+\frac{1}%
{2}\hat{P}^{i}H_{\beta}^{i}\mathbf{u}^{2}\alpha^{\beta}\\
& +iu_{\gamma}\varepsilon^{\gamma\beta\delta}\mathbf{\Sigma}^{\delta}%
u_{\gamma}\alpha^{\gamma}H_{\beta}^{i}\hat{P}^{i}-H_{\beta}^{i}\hat{P}%
^{i}u_{\gamma}\alpha^{\gamma}iu_{\gamma}\varepsilon^{\gamma\beta\delta
}\mathbf{\Sigma}^{\delta}\\
& +\frac{1}{2}iu_{\gamma}\varepsilon^{\gamma\beta\delta}\left(  \left[
\mathbf{\Sigma}^{\delta}H_{\beta}^{i}\left[  \hat{P}^{i},u_{\gamma}%
\alpha^{\gamma}\right]  +H_{\beta}^{i}\left[  \hat{P}^{i},u_{\gamma}%
\alpha^{\gamma}\right]  \mathbf{\Sigma}^{\delta}\right]  \right) \\
& =\frac{1}{2}\alpha^{\beta}\mathbf{u}^{2}H_{\beta}^{i}\hat{P}^{i}+\frac{1}%
{2}\hat{P}^{i}H_{\beta}^{i}\mathbf{u}^{2}\alpha^{\beta}\\
& +\left[  \left(  \mathbf{u}^{2}\Sigma^{\beta}-\left(  \mathbf{u\Sigma
}\right)  u^{\beta}\right)  H_{\beta}^{i}\hat{P}^{i}-H_{\beta}^{i}\hat{P}%
^{i}\left(  \mathbf{u}^{2}\Sigma^{\beta}-\left(  \mathbf{u\Sigma}\right)
u^{\beta}\right)  \right] \\
& +\frac{1}{2}iu_{\gamma}\varepsilon^{\gamma\beta\delta}\left(  \left[
\mathbf{\Sigma}^{\delta}H_{\beta}^{i}\left[  \hat{P}^{i},u_{\gamma}%
\alpha^{\gamma}\right]  +H_{\beta}^{i}\left[  \hat{P}^{i},u_{\gamma}%
\alpha^{\gamma}\right]  \mathbf{\Sigma}^{\delta}\right]  \right)
\end{align*}
Now, given that:
\begin{align*}
& iu_{\gamma}\varepsilon^{\gamma\beta\delta}\left(  \left[  \mathbf{\Sigma
}^{\delta}H_{\beta}^{i}\left[  \hat{P}^{i},u_{\gamma}\alpha^{\gamma}\right]
+H_{\beta}^{i}\left[  \hat{P}^{i},u_{\gamma}\alpha^{\gamma}\right]
\mathbf{\Sigma}^{\delta}\right]  \right) \\
& =-\hbar u_{\gamma}\varepsilon^{\gamma\kappa\delta}\left(  \left[
\mathbf{\Sigma}^{\delta}H_{\kappa}^{i}\left(  -\nabla_{i}u_{\gamma}%
\alpha^{\gamma}+\frac{\hat{\Gamma}_{i}^{\varrho\beta}}{2}\mathbf{\Sigma
}^{\beta}u_{\varrho}\right)  +H_{\kappa}^{i}\left(  -\nabla_{i}u_{\gamma
}\alpha^{\gamma}+\frac{\hat{\Gamma}_{i}^{\varrho\beta}}{2}\mathbf{\Sigma
}^{\beta}u_{\varrho}\right)  \mathbf{\Sigma}^{\delta}\right]  \right) \\
& =-\hbar u_{\gamma}\varepsilon^{\gamma\kappa\delta}H_{\kappa}^{i}\left(
-2J\nabla_{i}u_{\delta}+\hat{\Gamma}_{i}^{\varrho\delta}u_{\varrho}\right)
\end{align*}
one thus has,
\begin{align*}
u_{\beta}\alpha^{\beta}\frac{\left(  \hat{H}^{\prime}+\hat{H}^{\prime
+}\right)  }{2}u_{\beta}\alpha^{\beta}  & =\frac{1}{2}\alpha^{\beta}%
\mathbf{u}^{2}\hat{P}_{\beta}+\frac{1}{2}\hat{P}_{\beta}\mathbf{u}^{2}%
\alpha^{\beta}-\frac{1}{2}g_{00}h_{0}^{0}\hbar u_{\gamma}\varepsilon
^{\gamma\kappa\delta}H_{\kappa}^{i}\left(  -2J\nabla_{i}u_{\delta}+\hat
{\Gamma}_{i}^{\varrho\delta}u_{\varrho}\right) \\
& =\frac{1}{2}\alpha^{\beta}\mathbf{u}^{2}\hat{P}_{\beta}+\frac{1}{2}\hat
{P}_{\beta}\mathbf{u}^{2}\alpha^{\beta}-\frac{1}{2}g_{00}h_{0}^{0}\hbar
u_{\gamma}\varepsilon^{\gamma\kappa\delta}H_{\kappa}^{i}\left(  -2J\nabla
_{i}u_{\delta}\right)
\end{align*}
since, by an argument already used, $u_{\gamma}\varepsilon^{\gamma\kappa
\delta}H_{\kappa}^{i}\hat{\Gamma}_{i}^{\varrho\delta}u_{\varrho}\propto\left(
\mathbf{h}^{0}\times\mathbf{h}^{i}\right)  _{\delta}\hat{\Gamma}_{i}%
^{\varrho\delta}u_{\varrho}=0$.

To complete the computation of $u_{\beta}\alpha^{\beta}\frac{\left(
H+H^{+}\right)  }{2}u_{\beta}\alpha^{\beta}$, we need to calculate the
following contribution:
\begin{align*}
& u_{\gamma}\alpha^{\gamma}\left(  g_{00}h_{0}^{0}\beta m+\frac{1}{2}%
g_{00}g^{0i}P_{i}+\frac{1}{2}P_{i}g_{00}g^{0i}+\left(  \hat{\Gamma}_{\gamma
}^{0}-\hbar\nabla_{R_{i}}\left(  g_{00}\left(  \mathbf{h}^{0}\times
\mathbf{h}^{i}\right)  \right)  _{\gamma}\right)  \mathbf{\Sigma}^{\gamma
}\right)  u_{\gamma}\alpha^{\gamma}\\
& =-g_{00}h_{0}^{0}\beta\mathbf{u}^{2}m+\frac{1}{2}g_{00}g^{0i}\mathbf{u}%
^{2}P_{i}+\frac{1}{2}\mathbf{u}^{2}P_{i}g_{00}g^{0i}+\mathbf{u}^{2}\left(
\hat{\Gamma}_{\gamma}^{0}-\hbar\nabla_{R_{i}}\left(  g_{00}\left(
\mathbf{h}^{0}\times\mathbf{h}^{i}\right)  _{\gamma}\right)  \right)
\mathbf{\Sigma}^{\gamma}\\
& -2\mathbf{u}^{2}\left(  \hat{\Gamma}_{\gamma}^{0}-\hbar\nabla_{R_{i}}\left(
g_{00}\left(  \mathbf{h}^{0}\times\mathbf{h}^{i}\right)  _{\gamma}\right)
\right)  \Sigma^{\gamma}+\mathbf{u.\Sigma}\left(  \hat{\Gamma}_{\gamma}%
^{0}-\hbar\nabla_{R_{i}}\left(  g_{00}\left(  \mathbf{h}^{0}\times
\mathbf{h}^{i}\right)  _{\gamma}\right)  \right)  u_{\gamma}\\
& =-g_{00}h_{0}^{0}\beta\mathbf{u}^{2}m+\frac{1}{2}g_{00}g^{0i}\mathbf{u}%
^{2}P_{i}+\frac{1}{2}\mathbf{u}^{2}P_{i}g_{00}g^{0i}+\mathbf{u.\Sigma}\left(
\hat{\Gamma}_{\gamma}^{0}-\hbar\nabla_{R_{i}}\left(  g_{00}\left(
\mathbf{h}^{0}\times\mathbf{h}^{i}\right)  _{\gamma}\right)  \right)
u_{\gamma}\\
& -\mathbf{u}^{2}\left(  \hat{\Gamma}_{\gamma}^{0}-\hbar\nabla_{R_{i}}\left(
g_{00}\left(  \mathbf{h}^{0}\times\mathbf{h}^{i}\right)  _{\gamma}\right)
\right)  \mathbf{\Sigma}^{\gamma}%
\end{align*}
Now, we turn our attention toward the second contribution in Eq. $\left(
\ref{dev}\right)  $, namely $\frac{-\left[  \hat{H},u_{\beta}\alpha^{\beta
}\right]  +\left[  \hat{H}^{+},u_{\beta}\alpha^{\beta}\right]  }{2}$. To do
so, \ and since
\[
-\frac{\left[  \hat{H},u_{\beta}\alpha^{\beta}\right]  +\left[  \hat{H}%
^{+},u_{\beta}\alpha^{\beta}\right]  }{2}=-\frac{1}{2}\left[  \hat{H}-\hat
{H}^{+},u_{\beta}\alpha^{\beta}\right]
\]
we only need to concentrate on the anti hermitian part of $H$. Given that,%

\begin{align*}
\hat{H}  & =\alpha^{\beta}H_{\beta}^{i}\left(  P_{i}+\hbar\varepsilon
_{\varrho\beta\gamma}\frac{\hat{\Gamma}_{i}^{\varrho\beta}}{4}\mathbf{\Sigma
}^{\gamma}+i\hbar\frac{\Gamma_{i}^{0\gamma}}{4}\alpha^{\gamma}\right)
+g_{00}g^{0i}i\hbar\frac{\Gamma_{i}^{0\gamma}}{4}\alpha^{\gamma}\\
& +ig_{00}\left(  h_{\delta}^{0}h_{\eta}^{i}\varepsilon^{\delta\eta\kappa
}\right)  \mathbf{\Sigma}_{\kappa}(P_{i}+\hbar\varepsilon_{\varrho\beta\gamma
}\frac{\Gamma_{i}^{\varrho\beta}}{4}\mathbf{\Sigma}^{\gamma})+i\frac{\hbar}%
{4}\Gamma_{0}^{0\beta}\alpha^{\beta}%
\end{align*}
we can write that
\begin{align*}
\frac{\hat{H}-\hat{H}^{+}}{2}  & =i\frac{\hbar}{2}\nabla_{i}\left(
\alpha^{\beta}H_{\beta}^{i}\right)  +iH_{\beta}^{i}\hbar\frac{\Gamma
_{i}^{0\beta}}{4}+g_{00}g^{0i}i\hbar\frac{\Gamma_{i}^{0\beta}}{4}\alpha
^{\beta}-i\frac{\hbar}{2}H_{\varrho}^{i}\alpha_{\beta}\hat{\Gamma}_{i}%
^{\beta\varrho}\\
& +\frac{i}{2}g_{00}\left(  h_{\delta}^{0}h_{\eta}^{i}\varepsilon^{\delta
\eta\kappa}\right)  \mathbf{\Sigma}_{\kappa}P_{i}+\frac{i}{2}P_{i}%
g_{00}\left(  h_{\delta}^{0}h_{\eta}^{i}\varepsilon^{\delta\eta\kappa}\right)
\mathbf{\Sigma}_{\kappa}+ig_{00}\left(  h_{\delta}^{0}h_{\eta}^{i}\right)
\hbar\frac{\Gamma_{i}^{\delta\eta}}{2}+i\frac{\hbar}{4}\Gamma_{0}^{0\beta
}\alpha^{\beta}%
\end{align*}
and
\begin{align*}
-\frac{\left[  \hat{H},u_{\beta}\alpha^{\beta}\right]  +\left[  \hat{H}%
^{+},u_{\beta}\alpha^{\beta}\right]  }{2}  & =-\hbar\left(  \nabla_{i}\left(
H_{\beta}^{i}\right)  +g_{00}g^{0i}\frac{\Gamma_{i}^{0\beta}}{2}+\frac
{\Gamma_{0}^{0\beta}}{2}-H_{\varrho}^{i}\hat{\Gamma}_{i}^{\beta\varrho
}\right)  \varepsilon^{\beta\gamma\kappa}u_{\gamma}\Sigma_{\kappa}\\
& +g_{00}\left(  h_{\delta}^{0}h_{\eta}^{i}-h_{\eta}^{0}h_{\delta}^{i}\right)
u_{\delta}\alpha_{\eta}P_{i}+P_{i}g_{00}\left(  h_{\delta}^{0}h_{\eta}%
^{i}-h_{\eta}^{0}h_{\delta}^{i}\right)  u_{\delta}\alpha_{\eta}\\
& -g_{00}\hbar\left(  h_{\delta}^{0}h_{\eta}^{i}\varepsilon^{\delta\eta\kappa
}\right)  J\nabla_{i}u_{\kappa}%
\end{align*}
Ultimately, we need the third contribution to Eq. $\left(  \ref{dev}\right)  $:%

\begin{align*}
& -i\frac{\hbar}{2}\left[  u_{\beta}\alpha^{\beta},\left(  \mathbf{\nabla
}_{\mathbf{P}}\hat{H}\right)  .\mathbf{\nabla}_{\mathbf{R}}\left(  \frac
{1}{f\left(  1-u_{\beta}u^{\beta}\right)  }\right)  \right] \\
& =i\frac{\hbar}{2}\left[  u_{\beta}\alpha^{\beta},\alpha^{\beta}H_{\beta}%
^{i}.\left(  \frac{\nabla_{R_{i}}\left(  f\left(  1-u_{\beta}u^{\beta}\right)
\right)  }{\left[  f\left(  1-u_{\beta}u^{\beta}\right)  \right]  ^{2}%
}\right)  \right] \\
& =-\hbar\mathbf{\Sigma.}\left(  \mathbf{u\times H}^{i}\right)  \left(
\frac{\nabla_{R_{i}}\left(  f\left(  1-\mathbf{u}^{2}\right)  \right)
}{\left[  f\left(  1-\mathbf{u}^{2}\right)  \right]  ^{2}}\right)
\end{align*}
We can now gather all these terms to obtain the expression of $U^{\frac{1}{2}%
}\hat{H}U^{-\frac{1}{2}}$.

Define, as in text, the vectors $\mathbf{\tilde{\Gamma}}_{i}^{\delta}$,
$\mathbf{\Gamma}_{i}^{0}$ by:
\[
\left(  \mathbf{\tilde{\Gamma}}_{i}^{\delta}\right)  ^{\beta}=\tilde{\Gamma
}_{i}^{\delta\beta},\left(  \mathbf{\Gamma}_{i}^{0}\right)  ^{\beta}%
=\Gamma_{i}^{0\beta},\left(  \mathbf{\Gamma}_{0}^{0}\right)  ^{\beta}%
=\Gamma_{0}^{0\beta}
\]
and $\tilde{H}_{\beta}^{i}$:
\[
\tilde{H}_{\beta}^{i}=H_{\beta}^{i}+\frac{2g_{00}\left(  h_{\delta}%
^{0}h_{\beta}^{i}-h_{\beta}^{0}h_{\delta}^{i}\right)  u_{\delta}}{\left(
1-\mathbf{u}^{2}\right)  }
\]%
\[
\tilde{\Gamma}_{i}^{\varrho\beta}=\left(  \tilde{H}^{-1}\right)  _{i}^{\eta
}H_{\eta}^{j}\hat{\Gamma}_{j}^{\varrho\beta}
\]
Ultimately, reintroducing the factors $\frac{1}{\left(  1-\mathbf{u}%
^{2}\right)  }$ when needed in Eq. $\left(  \ref{dev}\right)  $, $U^{\frac
{1}{2}}\hat{H}U^{-\frac{1}{2}}$ can be written in a compact form :
\begin{align*}
U^{\frac{1}{2}}\hat{H}U^{-\frac{1}{2}}  & =\frac{1}{2}\alpha^{\beta}\tilde
{H}_{\beta}^{i}\left(  P_{i}+\hbar\varepsilon_{\varrho\beta\gamma}\frac
{\tilde{\Gamma}_{i}^{\varrho\beta}}{4}\Sigma^{\gamma}\right)  +\frac{1}%
{2}\left(  P_{i}+\hbar\varepsilon_{\varrho\beta\gamma}\frac{\tilde{\Gamma}%
_{i}^{\varrho\beta}}{4}\Sigma^{\gamma}\right)  \tilde{H}_{\beta}^{i}%
\alpha^{\beta}\\
& +g_{00}h_{0}^{0}\frac{\left(  1+\mathbf{u}^{2}\right)  }{\left(
1-\mathbf{u}^{2}\right)  }\beta m+\frac{1}{2}g_{00}g^{0i}P_{i}+\frac{1}%
{2}P_{i}g_{00}g^{0i}\\
& +\left(  \tilde{\Gamma}_{\gamma}^{0}+\mathbf{\Gamma}^{e}\right)
.\mathbf{\Sigma}-\hbar\frac{\left(  g_{00}\left(  \mathbf{h}^{0}%
\times\mathbf{h}^{i}\right)  +g_{00}h_{0}^{0}\mathbf{u\times H}^{i}\right)
}{\left(  1-\mathbf{u}^{2}\right)  }.\left(  \nabla_{i}\mathbf{u}\right)  J
\end{align*}
where we have defined $\tilde{\Gamma}_{\gamma}^{0}$ by :
\[
\tilde{\Gamma}_{\gamma}^{0}=\frac{\left(  1+\mathbf{u}^{2}\right)
\delta_{\gamma}^{\eta}-u^{\eta}u_{\gamma}}{\left(  1-\mathbf{u}^{2}\right)
}\hat{\Gamma}_{\eta}^{0}+\hbar\left(  \left(  H_{\delta}^{i}\mathbf{\hat
{\Gamma}}_{i}^{\delta}\mathbf{+}\frac{\mathbf{\Gamma}_{0}^{0}}{2}%
\mathbf{-}g_{00}g^{0i}\frac{\mathbf{\Gamma}_{i}^{0}}{2}\right)  \times
\mathbf{u}\right)  _{\gamma}
\]%
\[
\mathbf{\Gamma}_{\gamma}^{e}=-\hbar\left(  \mathbf{u\times H}^{i}\right)
_{\gamma}\left(  \frac{\nabla_{R_{i}}\left(  f\left(  1-\mathbf{u}^{2}\right)
\right)  }{f^{2}\left(  1-\mathbf{u}^{2}\right)  ^{3}}\right)  +\hbar
\frac{\left(  \mathbf{u\times}\nabla_{i}\left(  \mathbf{H}^{i}\right)
\right)  _{\gamma}}{\left(  1-\mathbf{u}^{2}\right)  }-\hbar\frac{\left(
1+\mathbf{u}^{2}\right)  \delta_{\gamma}^{\eta}-u^{\eta}u_{\gamma}}{\left(
1-\mathbf{u}^{2}\right)  }\nabla_{R_{i}}\left(  g_{00}\left(  \mathbf{h}%
^{0}\times\mathbf{h}^{i}\right)  _{\eta}\right)
\]

\subsection{Non Hermitian contributions of the time evolution operator}

We compute here the contributions to the diagonalization due to the
non-hermitian term $-i\hbar U^{\frac{1}{2}}\left(  t\right)  \frac{\partial
}{\partial t}U^{-\frac{1}{2}}\left(  t\right)  $ . First we obain%
\begin{equation}
-i\hbar U^{\frac{1}{2}}\left(  t\right)  \frac{\partial}{\partial t}%
U^{-\frac{1}{2}}\left(  t\right)  =i\hbar\frac{\frac{\partial}{\partial
t}\left[  f\left(  1-\mathbf{u}^{2}\right)  \right]  }{f\left(  1-\mathbf{u}%
^{2}\right)  }+i\hbar\frac{\left(  1+u_{\beta}\frac{\partial}{\partial
t}u_{\beta}\right)  }{\left(  1-\mathbf{u}^{2}\right)  }+i\hbar\frac
{\frac{\partial}{\partial t}u_{\beta}\alpha^{\beta}}{\left(  1-\mathbf{u}%
^{2}\right)  }-\hbar\frac{\left(  \mathbf{u\times}\frac{\partial}{\partial
t}\mathbf{u}\right)  .\mathbf{\Sigma}}{\left(  1-\mathbf{u}^{2}\right)
}\label{anx}%
\end{equation}

and then, the contributions of the terms in Eq.$\left(  \ref{anx}\right)  $ to
the diagonalization are computed by applying the transformation matrix and
then projecting on the diagonal blocks. We are left with:
\begin{align*}
U\left(  \mathbf{\tilde{P}}\right)  \left[  -i\hbar U^{\frac{1}{2}}\left(
t\right)  \frac{\partial}{\partial t}U^{-\frac{1}{2}}\left(  t\right)
\right]  U\left(  \mathbf{\tilde{P}}\right)  ^{+}  & =i\hbar\frac
{\frac{\partial}{\partial t}\left[  f\left(  1-\mathbf{u}^{2}\right)  \right]
}{f\left(  1-\mathbf{u}^{2}\right)  }+i\hbar\frac{\left(  1+\mathbf{u.}%
\frac{\partial}{\partial t}\mathbf{u}\right)  }{\left(  1-\mathbf{u}%
^{2}\right)  }+i\hbar c\beta\frac{\frac{\partial}{\partial t}\mathbf{u}%
.\mathbf{\tilde{P}}}{\left(  1-\mathbf{u}^{2}\right)  E_{0}}\\
& -\frac{\hbar c^{2}\left(  \mathbf{\tilde{P}\times}\left(  \mathbf{\tilde
{P}\times}\left(  \mathbf{u\times}\frac{\partial}{\partial t}\mathbf{u}%
\right)  \right)  .\mathbf{\Sigma}\right)  }{\left(  1-\mathbf{u}^{2}\right)
E_{0}\left(  E_{0}+\tilde{m}\right)  }%
\end{align*}

\bigskip

\bigskip
\end{document}